\documentclass[aps,english,prl,amsmath,amsfonts,amssymb,superscriptaddress,twocolumn,showpacs,citeautoscript]{revtex4-1}
\usepackage[T1]{fontenc}
\usepackage[latin9]{inputenc}
\usepackage{amsmath}
\usepackage{amssymb}
\usepackage{wasysym}
\usepackage{esint}
\usepackage{graphicx}
\usepackage{times}
\usepackage{nicefrac}
\usepackage{appendix}
\usepackage{textcase}
\usepackage{subfigure}
\usepackage{empheq}
\usepackage{color}
\usepackage{leftidx}
\usepackage{makecell}
\usepackage{stackrel}
\makeatletter
\@ifundefined{textcolor}{}
{%
  \definecolor{BLACK}{gray}{0}
  \definecolor{WHITE}{gray}{1}
  \definecolor{RED}{rgb}{1,0,0}
  \definecolor{GREEN}{rgb}{0,1,0}
  \definecolor{BLUE}{rgb}{0,0,1}
  \definecolor{CYAN}{cmyk}{1,0,0,0}
  \definecolor{MAGENTA}{cmyk}{0,1,0,0}
  \definecolor{YELLOW}{cmyk}{0,0,1,0}
}

\renewcommand{\eqref}[1]{(\ref{eq:#1})}

\newcommand{\figref}[1]{Fig.~\ref{fig:#1}}

\newcommand{\appref}[1]{Appendix~\ref{sec:#1}}

\newcommand{\Secref}[1]{Section~\ref{sec:#1}}

\renewcommand{\vec}[1]{\mathbf{#1}}
\newcommand{\ket}[1]{|#1\rangle}
\newcommand{\bra}[1]{\langle#1|}

\newcommand{\bracket}[1]{\langle#1\rangle}

\newcommand{\im}{\operatorname{i}}
\renewcommand{\Re}{\operatorname{Re}}
\renewcommand{\Im}{\operatorname{Im}}
\newcommand{\asym}{\operatorname{asym}}
\newcommand{\sym}{\operatorname{sym}}
\newcommand{\erfc}{\operatorname{erfc}}
\newcommand{\trace}{\operatorname{Tr}}
\newcommand{\tr}{\operatorname{Tr}}

\newcommand{\II}{\mathbb{I}}
\newcommand{\VV}{\mathbb{V}}
\newcommand{\GG}{\mathbb{G}}
\newcommand{\PP}{\mathbb{P}}
\newcommand{\TT}{\mathbb{T}}
\newcommand{\WW}{\mathbb{W}}

\begin{document}
\title{Fluctuational Electrodynamics in Atomic and Macroscopic
  Systems: van der Waals Interactions and Radiative Heat Transfer}

\author{Prashanth S. Venkataram}
\affiliation{Department of Electrical Engineering, Princeton
  University, Princeton, New Jersey 08544, USA}

\author{Jan Hermann}
\affiliation{Department of Mathematics and Computer Science, FU
  Berlin, Arnimallee 6, 141 95 Berlin, Germany}

\author{Alexandre Tkatchenko}
\affiliation{Physics and Materials Science Research Unit, University
  of Luxembourg, L-1511 Luxembourg}

\author{Alejandro W. Rodriguez}
\affiliation{Department of Electrical Engineering, Princeton
  University, Princeton, New Jersey 08544, USA}

\date{\today}

\begin{abstract}
  We present an approach to describing fluctuational electrodynamic
  interactions, particularly van der Waals (vdW) interactions as well
  as radiative heat transfer (RHT), between material bodies of
  potentially vastly different length scales, allowing for going
  between atomistic and continuum treatments of the response of each
  of these bodies as desired. Any local continuum description of
  electromagnetic response is compatible with our approach, while
  atomistic descriptions in our approach are based on effective
  electronic and nuclear oscillator degrees of freedom, encapsulating
  dissipation, short-range electronic correlations, and collective
  nuclear vibrations (phonons). While our previous works using this
  approach have focused on presenting novel results, this work focuses
  on the derivations underlying these methods. First, we show how the
  distinction between ``atomic'' and ``macroscopic'' bodies is
  ultimately somewhat arbitrary, as formulas for vdW free energies and
  radiative heat transfer look very similar regardless of how the
  distinction is drawn. Next, we demonstrate that the atomistic
  description of material response in our approach yields
  electromagnetic interaction matrix elements which are expressed in
  terms of analytical formulas for compact bodies or semianalytical
  formulas based on Ewald summation for periodic media; we use this to
  compute vdW interaction free energies as well as RHT powers among
  small biological molecules in the presence of a metallic plate as
  well as between parallel graphene sheets in vacuum, showing strong
  deviations from conventional macroscopic theories due to the
  confluence of geometry, phonons, and electromagnetic retardation
  effects. Finally, we propose formulas for efficient computation of
  fluctuational electrodynamic interactions among material bodies in
  which those that are treated atomistically as well as those treated
  through continuum methods may have arbitrary shapes, extending
  previous surface-integral techniques.
\end{abstract}

\maketitle


\section{Introduction}



Quantum and thermal fluctuations in electromagnetic (EM) fields are
modified in the presence of polarizable objects. In thermal
equilibrium, these fluctuating fields can transfer momentum, effecting
van der Waals (vdW) or (synonymously) Casimir interactions, while out
of thermal equilibrium, they can transfer energy, effecting thermal
radiation and heat transfer between bodies. vdW interactions are of
particular importance to molecular and low-dimensional structures both
large and small, determining binding energies, stable conformations of
polymorphic noncovalent crystals, and mechanical
properties~\cite{ReillyPRL2014, HojaCMS2017, TkatchenkoJCP2013,
  TkatchenkoJCP2013, DiStasioJPCM2014, TkatchenkoADFM2014}. Recent
studies of vdW interactions in molecular materials have illustrated
the importance of modeling vdW interactions beyond the regime of
pairwise additivity~\cite{ReillyCS2015, AmbrosettiSCIENCE2016,
  AmbrosettiJCP2014, TkatchenkoPRL2012}, which is valid only for
isolated atoms/small molecules or (equivalently) sufficiently dilute
bulk media, though these works have only considered distance regimes
where the EM field response may be taken as the Coulomb potential in
the electrostatic limit. On the other hand, theoretical studies of
Casimir interactions among macroscopic bodies~\cite{Johnson2011,
  RahiPRD2009, RodriguezPRL2007, RahiPRA2008, LevinPRL2010} have
demonstrated nonmonotonic and repulsive forces among conducting
objects even in vacuum at much larger distance scales where the speed
of light (EM retardation) matters, but such continuum treatments are
generally restricted to size and distance regimes large enough that
continuous local empirically-fitted dispersive susceptibilities
accurately model the polarization response, so they are unable to
accurately capture the atomistic nature and nonlocality (spatial
dispersion) of the response of smaller molecular systems. Meanwhile,
theoretical descriptions of radiative heat transfer (RHT) have been
largely restricted to macroscopic bodies modeled with continuum local
susceptibilities~\cite{MessinaPRB2017, JinOE17, RodriguezPRL2011,
  OteyJQSRT2014}, demonstrating large enhancements as well as
suppression factors arising from the tunneling of surface waves at
short body separations, as compared to the predictions of the Planck
blackbody law; in contrast to the case of vdW interactions, only a
handful of investigations of RHT have focused on atomistic
structures~\cite{PendryPRB2016, ChiloyanNATURE2015, PendryJPCM1999,
  DominguesPRL2005}. Comparatively more work has been pursued in the
context of conductive heat transport by electrons and
phonons~\cite{TianPRB2012, TianPRB2014, DharJSP2006, MingoPRB2003},
but even there, existing models tend to be fully atomistic and
therefore restricted to treatments of small molecules or simple
geometries with a high degree of (e.g. translational or rotational)
symmetry. Since phonons and plasmons arise from and are influenced by
EM interactions, respectively, fundamental questions remain
surrounding the transition between radiative and conductive heat
transfer at subnanometric gaps~\cite{KimNATURE2015,
  KloppstechNATURE2017, CuiNATURE2017, ChiloyanNATURE2015}.

\begin{figure*}[t!]
  \centering
  \includegraphics[width=0.95\textwidth]{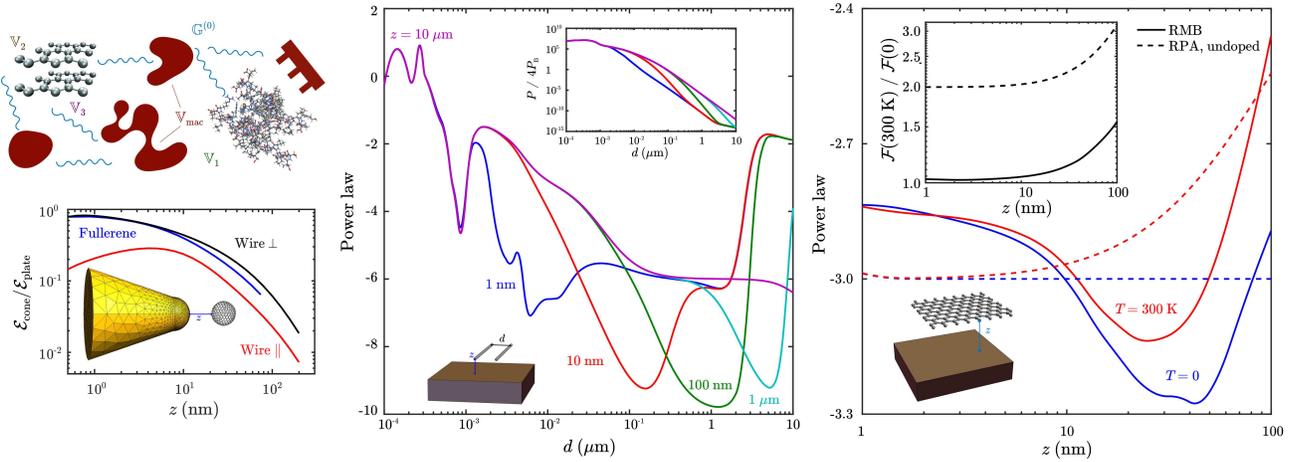}
  \caption{\textbf{Diverse sample of past results in the RMB
      framework}. Top-left, adapted from~\cite{VenkataramPRL2017}:
    schematic of general systems that may be investigated in the RMB
    framework, involving atom-scale and macroscopic
    bodies. Bottom-left, adapted from~\cite{VenkataramPRL2017}: vdW
    interaction energy (at $T = 0$) of a $\mathrm{C}_{500}$-fullerene
    or 250 atom-long carbyne wire in the parallel or perpendicular
    orientations with respect to a gold conical tip at distance $z$,
    relative to the interaction energy with respect to a semi-infinite
    planar gold slab at the same $z$. Middle, adapted
    from~\cite{VenkataramPRL2018}: exponent (power law) of RHT power
    with respect to distance $d$ separating two parallel 500 atom-long
    carbyne wires, each at distance $z$ from a PEC plane. Middle
    inset: RHT power at each $z$, normalized to 4 times the equivalent
    blackbody emission power, as functions of $d$. Right, adapted
    from~\cite{VenkataramSCIADV2019}: exponent (power law) of vdW
    interaction free energy at zero or room temperatures between an
    infinite graphene sheet and a parallel PEC plane with respect to
    the separation $z$, comparing RMB predictions (solid) to continuum
    RPA predictions without doping (dashed). Right inset: ratios of
    vdW interaction free energies at room to zero temperatures in RMB
    or RPA frameworks.}
  \label{fig:pastwork}
\end{figure*}

Recently, we proposed a theoretical approach that conjoins atomistic
treatments of molecular and low-dimensional structures with continuum
treatments of macroscopic bodies in the context of fluctuational
electrodynamics (FED) to enable description of EM fluctuation effects
over a wide range of distance and geometric scales (from atom- to
micron-scale gaps and from molecular to macroscopic media), including
situations in which continuum approximations fail for a subset of the
interacting bodies but not for others. We have called this the
retarded many-body (RMB) framework of mesoscale FED. Our work has
illustrated the importance of retardation effects in small molecular
systems (where they are typically assumed to be negligible) and of
geometry in determining the impact of collective, long-range EM
fluctuations (i.e. polaritons) that cannot be appropriately captured by
pairwise--additive approximations~\cite{VenkataramPRL2017}.
Furthermore, we have shown that phonons in molecular structures can
delocalize the polarization response of large molecules, leading for
instance to nontrivial corrections to vdW interactions at room
temperature relative to purely quantum
fluctuations~\cite{VenkataramSCIADV2019}. Similar consideration of
phonons, nonlocal response, and long-range EM effects play a critical
role in describing heat transfer among proximate molecules, with
ab-initio atomistic modeling of the molecular response of materials
enabling accurate descriptions of the transition from radiative to
conductive heat transfer within the same unified theoretical
framework~\cite{VenkataramPRL2018}. A sample of results from these
papers is in~\figref{pastwork}.


Early experiments on vdW interactions and RHT focused on measuring
these phenomena in simple planar or spherical geometries validating
predictions from continuum models of material
response~\cite{ZhaoSCIENCE2019, MundayNATURE2009, BanishevPRB2013,
  SushkovNATURE2011, KlimchitskayaPRB2015, GarrettPRL2018,
  CahillAPR2014, CuevasACS2018, HargreavesPLA1969, HuAPL2008,
  RousseauNATURE2009, OttensPRL2011, KralikRSI2011, ShenAPL2012},
while later experiments within the continuum domain have gone beyond
such simple geometries~\cite{ChenPRL2002, GreffetNATURE2002,
  StGelaisNATURE2016, TangNATURE2017, InoueNANOLETT2019,
  SongNATURENANO2015}. More recent experiments in the atomistic and
continuum domains of FED have begun to emerge, probing the edges of
the regimes of validity of prior theoretical treatments, suggesting
the need for new theoretical frameworks to better treat such
multiscale problems. These experiments include measurements of vdW
forces between organic molecules, macromolecular arrays, or
single-layer sheets, and planar metallic or dielectric substrates
without retardation~\cite{WagnerNATURE2014, LoskillJRSI2013,
  TsoiACSNANO2014}, as well as between nanoparticles and
nanotubes~\cite{RanceACSNANO2010, SilveraBatistaSCIENCE2015}, that
explore situations beyond the pairwise additive regime, measurements
of Casimir--Polder forces on ground-state and Rydberg atoms,
molecules, and Bose--Einstein condensates near planar substrates and
gratings~\cite{BenderPRX2014, Intravaia2011, DeKieviet2011,
  Buhmann2012I, Buhmann2012II} where EM retardation is relevant,
measurements of near-field RHT between metallic tips and substrates at
nanometric gaps~\cite{KimNATURE2015, KloppstechNATURE2017,
  CuiNATURE2017, ChiloyanNATURE2015}, and observations of thermal
conductances in single-molecule junctions~\cite{CuiSCIENCE2017,
  CuiNATURE2019}. Such experiments are relevant to the engineering and
operational understanding of molecular-scale
devices~\cite{CuiSCIENCE2017, KlocknerPRB2016, ZouNATURE2013,
  AsanoSCIADV2016, GuhaNANOLETT2012}, heat management in electronic
and thermophotovoltaic devices~\cite{LenertNATURENANO2014,
  SongNATURENANO2015, KaralisSR2016, MessinaNATURE2013}, and
manipulation of living cells and nanoparticles used in
nanomedicine~\cite{ChenPNAS2003, ChattopadhyayaCM2017, WoodsRMP2015,
  ReillyPRL2014, NerngchamnongNATURE2013}, among other
applications. Accurate explanation of all of these experimental
results at many different length scales will require consideration of
the interplay of phonons, retarded EM response, and complex geometric
effects at the mesoscale, and suggests that our RMB framework may be
well-suited to answer such questions.


This paper accompanies a computational code which has been published
as open source for others to use and extend, so the main goal of this
paper is to provide rigorous derivations of the formulas underlying
this code as well as our previous works~\cite{VenkataramPRL2017,
  VenkataramPRL2018, VenkataramSCIADV2019}. In particular, we give
detailed derivations of the most general formulation of mesoscopic
FED, and show how our RMB approach combines sophisticated scattering
and electronic calculation techniques in regimes where accurate
atomic-scale descriptions of response are required for molecules while
continuum permittivity models suffice for larger bulk objects can be
exploited. Computational efficiency demands fast calculation of system
matrices representing scattering among microscopic degrees of freedom
(expanded in a basis of Gaussian functions): these matrices are found
to greatly simplify into semi-analytical formulas involving Gaussian
integrals in general, as well as Ewald summations in periodic media,
thereby speeding up matrix assembly, so we provide detailed
derivations of those formulas without assuming the absence of
retardation or the validity of point dipolar approximations. The
generality of our method allows for easy extension to other
mesoscale EM phenomena of interest, including deterministic phenomena
like absorption or scattering~\cite{ReidIEEE2015, PolimeridisIEEE2015}
and fluctuational phenomena like
fluorescence~\cite{PolimeridisPRB2015}, and while we do not focus on
such phenomena in this paper nor implement such computational routines
in our code, the open source nature of our code lowers barriers to
pursuing this line of work in the future. Additionally, our method is
general enough to consider macroscopic environments of arbitrary
geometries and material properties, but our code and prior
works~\cite{VenkataramPRL2017, VenkataramPRL2018,
  VenkataramSCIADV2019} have almost exclusively focused on idealized
perfect electrically conducting (PEC) planes as the archetypal
macroscopic body or else have made further approximations involving
compact molecules in the presence of more realistic macroscopic
bodies. Thus, in this paper, we provide rigorous derivations of the
extension of our method to treat arbitrary compact molecular and
macroscopic bodies in conjunction with each other, relaxing those
assumptions about the macroscopic body being a PEC plane. These
derivations are based on the surface-integral formulation of Maxwell's
equations~\cite{ReidIEEE2013, RodriguezPRB2013, ReidPRA2013}, and
while we do not computationally implement these formulas, our existing
code as well as the code required for macroscopic
computations~\cite{SCUFF} are both free and open source, making such a
conjunction more feasible for future work.

Prior treatments of vdW interaction and RHT phenomena in macroscopic
bodies have generally been related under the rubric of FED, allowing
for exploitation of state-of-the-art classical computational EM
techniques. These include finite-difference~\cite{RodriguezPRA2007,
  RodriguezPRA2009, McCauleyPRA2010, LuoPRL2004, RodriguezPRL2011,
  OteyJQSRT2014}, spectral~\cite{MessinaPRB2013, MessinaPRB2017,
  LambrechtNJP2006, DominguesPRL2005} and
T-operator~\cite{EmigPRA2003, EmigPRL2001, RahiPRA2008,
  MaghrebiPNAS2011, KrugerPRB2012, RahiPRD2009}, surface
integral~\cite{ReidIEEE2013, RodriguezPRB2013, ReidPRA2013}, and
volume integral~\cite{PolimeridisPRB2015} methods. All of these
methods depend on primarily local empirical models for macroscopic
susceptibilities and typically treat macroscopic objects as having
hard boundaries, while accounting for EM retardation and scattering to
all orders without making approximations about the smoothness of the
object surfaces nor the diluteness of the media involved. Among these,
only finite-difference time domain methods can handle spatially
dispersive and potentially nonlinear polarizability response
functions, and can handle any object geometry equally well; however,
this comes at the cost of needing to discretize all of space and
needing to step through time to a sufficient extent to obtain
converged results, making this method computationally inefficient in
most cases. The other methods are frequency domain methods, which
precludes consideration of material nonlinearity, but each has its
pros and cons beyond that; it is worth noting that all of the other
aforementioned methods besides finite-difference have the advantage of
discretizing only the degrees of freedom (DOFs) associated with each
object, without needing to discretize the space in between. Spectral
and T-operator methods converge most quickly for systems with
continuous translational or rotational symmetries, but for arbitrary
geometries, this convergence is drastically diminished. Surface and
volume integral approaches can be implemented using localized rather
than spectral basis functions, allowing for more efficient treatment
of arbitrary geometries; volume integral methods require
discretization of the volumes of every object, which is beneficial for
objects with spatially varying susceptibilities or temperature
gradients, while surface integral approaches are typically formulated
to work only with homogeneous materials in each object. All of these
methods can in principle handle linear spatially dispersive materials,
but the susceptibilities describing those spatially dispersive
materials tend to originate from phenomenological descriptions such as
the hydrodynamic model, which cannot easily be applied outside of the
simplest situations of spheres or planar substrates.

Atomistic treatments of heat transfer in molecular and larger systems
have come in a few different flavors. Molecular
dynamics~\cite{CuiJPCA2015, HenryPRL2008, EsfarjaniPRB2011,
  NoyaPRB2004} essentially solves Newton's equations of motion by
treating each atom or molecule as a particle interacting with other
such particles in a potential; this allows for conceptual simplicity,
and the use of the time domain allows for treating anharmonic effects,
but the treatment of noncovalent interactions among atoms and
molecules is typically via crude pairwise additive
approximations. Atomistic Green's function
methods~\cite{PendryPRB2016, ChiloyanNATURE2015, TianPRB2012,
  TianPRB2014, MingoPRB2003, DharJSP2006} in the frequency domain
allow for computation of heat transport by electrons or phonons, with
harmonic oscillator models of electronic and phononic coupling often
obtained from ab-initio calculations (while having the pitfall of
being unable to treat anharmonicity); however, treatment of radiative
heat transfer (i.e. via photons) is often neglected or heuristically
approximated in a pairwise fashion, and apart from that, these methods
tend to only be applicable to either small molecules or large bulk
media with no ability to effectively bridge between the two disparate
length scales. In the context of vdW interactions, recent
methods~\cite{TkatchenkoJCP2013, GobreNCOMMS2013, DiStasioJPCM2014,
  TkatchenkoADFM2014, AmbrosettiSCIENCE2016, ReillyCS2015,
  AmbrosettiJCP2014, TkatchenkoPRL2012, WoodsRMP2015, HermannCR2017,
  HojaCMS2017, ReillyPRL2014} have been successful by modeling
electrons in molecules as effective oscillators coupled by long-range
EM interactions, such that even though the underlying electronic model
does not capture the inherent delocalization of electrons in metallic
systems, long-range collective EM effects (i.e. polaritons) can be
properly captured at zero temperature. We have been able to extend
these methods~\cite{VenkataramPRL2017, VenkataramPRL2018,
  VenkataramSCIADV2019} to include the effects of EM retardation,
complex macroscopic geometries in the vicinity of molecules, and
phonons in molecular systems as treated in an ab-initio manner,
showing that especially at finite temperature, both vdW interactions
and heat transfer are strongly influenced by the nonlocal response
brought about in molecules by phonons, and that these interactions can
be significantly modified by the presence of other large macroscopic
bodies even when idealized as PEC planes.


This paper is organized as follows. After introducing matters of
notation \& convention and reviewing Maxwell's equations
in~\Secref{preliminaries}, we detail the general FED formulas for
mesoscopic systems involving molecular and macroscopic bodies
in~\Secref{generalsec}, starting with fully general formulas
irrespective of the continuum approximation, and then showing how
molecular and macroscopic DOFs can be partitioned. Following that,
in~\Secref{compact}, we briefly discuss how certain formulas simplify
for compact molecules interacting in a fixed macroscopic environment,
and the pitfalls therein, and demonstrate new results for interactions
among small biological molecules in the presence of a PEC plane. We
then use~\Secref{periodic} to provide rigorous detailed derivations of
the extensions of our general formulas to systems of infinite extent
with Bloch periodicity, and discuss vdW interactions and RHT between
two parallel graphene sheets in vacuum as an example of the
versatility of our method. Finally, we describe how to extend our
formalism to treat FED involving arbitrary compact molecules and
macroscopic bodies in the surface integral equation formulation of
Maxwell's equations in~\Secref{SIE}, and conclude
in~\Secref{conclusions}.

\section{Preliminaries} \label{sec:preliminaries}



A vector field $u_{i} (\vec{x})$ will be denoted as $\ket{\vec{u}}$;
we stress that the notation $\ket{\vec{u}}$ is a shorthand for a
semiclassical field, not a second-quantized state. The standard
conjugated inner product is defined as $\bracket{\vec{u}| \vec{v}} =
\sum_{i} \int u_{i}^{\star} (\vec{x}) v_{i} (\vec{x})~\mathrm{d}^{3}
x$. An operator $\mathbb{A}$ represents an integral kernel $A_{ij}
(\vec{x}, \vec{x}')$ such that $\ket{\vec{v}} = \mathbb{A}
\ket{\vec{u}}$ means $v_{i} (\vec{x}) = \sum_{j} \int A_{ij} (\vec{x},
\vec{x}') u_{j} (\vec{x}')~\mathrm{d}^{3} x'$; the identity operator,
denoted $\II$, is represented as $\delta^{3} (\vec{x} -
\vec{x}')\delta_{ij}$ in position space. The Hermitian adjoint
$\mathbb{A}^{\dagger}$ is defined in a basis-independent manner such
that $\bracket{\vec{u}| \mathbb{A}^{\dagger} \vec{v}} =
\bracket{\mathbb{A}\vec{u}| \vec{v}}$, so in position space,
$(\mathbb{A}^{\dagger})_{ij}(\vec{x}, \vec{x}') =
A^{\star}_{ji}(\vec{x}', \vec{x})$. In terms of this, the Hermitian
and anti-Hermitian parts of a square operator (whose domain and range
are the same), both of which are themselves Hermitian operators, are
respectively defined as $\sym(\mathbb{A}) = (\mathbb{A} +
\mathbb{A}^{\dagger})/2$ and $\asym(\mathbb{A}) = (\mathbb{A} -
\mathbb{A}^{\dagger})/(2\im)$, satisfying $\mathbb{A} =
\sym(\mathbb{A}) + \im\asym(\mathbb{A})$. The unconjugated transpose
$\mathbb{A}^{\top}$ is defined in position space such that
$(\mathbb{A}^{\top})_{ij}(\vec{x}, \vec{x}') = A_{ji}(\vec{x}',
\vec{x})$, and the complex conjugate $\mathbb{A}^{\star}$ in position
space satisfies $(\mathbb{A}^{\star})_{ij}(\vec{x}, \vec{x}') =
A^{\star}_{ij}(\vec{x}, \vec{x}')$; in terms of these, for square
operators, $\Re(\mathbb{A}) = (\mathbb{A} + \mathbb{A}^{\star})/2$ and
$\Im(\mathbb{A}) = (\mathbb{A} - \mathbb{A}^{\star})/(2\im)$. A
product of operators $\mathbb{A} \mathbb{B}$ represents the kernel
$\sum_{l} \int A_{il} (\vec{x}, \vec{x}'') B_{lj} (\vec{x}'',
\vec{x}')~\mathrm{d}^{3} x''$. Finally, the trace of an operator is
$\trace[\mathbb{A}] = \sum_{i} \int A_{ii} (\vec{x},
\vec{x})~\mathrm{d}^{3} x$ independent of basis. Once again, these are
all in the position space representation; other bases can be used as
convenient. Additionally, all quantities will be evaluated in the
frequency domain, so the dependence on the frequency $\omega$ will
generally be notationally suppressed for brevity and will only appear
explicitly as needed for clarity.

Maxwell's equations may be written in the frequency domain as
\begin{equation} \label{eq:Maxwellfield}
  \left[\nabla\times\left(\nabla\times\right) -
  \frac{\omega^{2}}{c^{2}} (\II + \VV)\right]\ket{\vec{E}} =
  \frac{\omega^{2}}{c^{2}} \ket{\vec{P}^{(0)}}
\end{equation}
describing the propagation of electric fields $\ket{\vec{E}}$ due to
free polarization sources $\ket{\vec{P}^{(0)}}$ in the presence of a
set of polarizable bodies, which could each be low-dimensional
atomistic structures or bulk media, with collective susceptibility
$\VV$. The susceptibility is related to the permittivity via
$\epsilon_{ij}(\vec{x}, \vec{x}') = \delta_{ij} \delta^{3} (\vec{x} -
\vec{x}') + V_{ij}(\vec{x}, \vec{x}')$, and relates the total
polarization density $\ket{\vec{P}}$ to the total electric field
$\ket{\vec{E}}$ via $\ket{\vec{P}} = \ket{\vec{P}^{(0)}} + \VV
\ket{\vec{E}}$. We focus solely on reciprocal media, in which the
relation $\VV = \VV^{\top}$ holds in position space (so
$V_{ij}(\vec{x}, \vec{x}') = V_{ji}(\vec{x}', \vec{x})$). Treating
inherently nonreciprocal materials like topological
insulators~\cite{HasanRMP2010, ZhuPRB2014, ZhuPRL2016, FuchsPRA2017}
(which could break reciprocity in the presence of an applied magnetic
field) or other intrinsic nonreciprocal magneto-optic
media~\cite{LevySR2017, BuhmannNJP2012} would require an extension of
this formalism beyond the scope of this paper, but metamaterials that
exhibit emergent nonreciprocal magneto-optic responses in an effective
medium framework~\cite{FanAOM2019, KhandekarNJP2019} can be treated
using our framework if the underlying materials obey reciprocity.


As these definitions are common in continuum EM theory but may be less
familiar in the context of quantum chemistry, we point out three items
of note. The first point is that the susceptibility in continuum EM
theory is often denoted $\chi$, especially for homogeneous, local, and
isotropic material responses, and may more generally be written as
$\chi_{ij}(\vec{x}, \vec{x}')$. This contrasts with conventions in
quantum chemistry, which define the charge density response as
$\chi(\vec{x}, \vec{x}') = \sum_{i,j} \frac{\partial}{\partial x_{i}}
\frac{\partial}{\partial x_{j}'} \alpha_{ij}(\vec{x}, \vec{x}')$ in
terms of the ``polarizability tensor'' $\alpha_{ij}(\vec{x},
\vec{x}')$, the latter of which is identical to our susceptibility
$V_{ij}(\vec{x}, \vec{x}')$~\cite{HermannCR2017}; strictly speaking,
the charge density response that corresponds to the susceptibility is
typically taken as the noninteracting charge density response. To
avoid confusion, we exclusively use the notation $\VV$ (and its
position space representation) for the susceptibility. Readers may
refer to~\appref{glossary} for more details.

The second point, related to the first, is that in quantum chemical
settings where retardation effects are unimportant, the induced
polarization density $\Delta \vec{P}(\vec{x})$ is of less interest
than the bound charge density $\rho_{\mathrm{B}} (\vec{x}) = -\nabla
\cdot \Delta \vec{P}(\vec{x})$, so the same physical bound charge
density can be reproduced by shifting $\Delta\vec{P}(\vec{x})$ by
$\nabla \times \vec{Q}(\vec{x})$ for an arbitrary gauge field
$\vec{Q}(\vec{x})$: this is a reflection of the fact that in the
absence of EM retardation, all electric fields are longitudinal and
irrotational, so the addition of incompressible (solenoidal) vector
fields cannot change the electrostatic properties of the
system. However, such gauge freedom in $\Delta \vec{P}(\vec{x})$ is
lost when retardation is important, as can be seen by rewriting
Maxwell's equations as $\left[\nabla\times\left(\nabla\times\right) -
  \frac{\omega^{2}}{c^{2}} \II\right]\ket{\vec{E}} =
\frac{\omega^{2}}{c^{2}} \ket{\vec{P}}$ where $\ket{\vec{P}} =
\ket{\vec{P}^{(0)}} + \ket{\Delta\vec{P}}$: the existence of
transverse radiative (electrodynamic) fields destroys such
electrostatic gauge invariance.

The third point, related to both of the prior two, is that it is more
common in quantum chemical treatments of vdW interactions to speak of
the free charge density $\rho^{(0)}(\vec{x})$ than the free
polarization field $\vec{P}^{(0)} (\vec{x})$ (and likewise the charge
density response instead of the susceptibility). In general, the two
are related in the frequency domain by $-\im\omega\nabla \cdot
\vec{P}^{(0)} (\vec{x}) - \im\omega\rho^{(0)}(\vec{x}) = 0$. It is
true that the free charge density becomes independent of the free
polarization field exactly at $\omega = 0$ (i.e. the static
regime). However, our formulations of vdW interactions and thermal
radiation depend on integrals over frequency in which the contribution
at exactly $\omega = 0$ is infinitesimal (and vanishes in the specific
case of thermal radiation). For this reason and also to fully account
for finite frequency effects (i.e. EM retardation) as well as
anisotropy, we consistently use the free polarization field
$\ket{\vec{P}^{(0)}}$ instead of $\rho^{(0)}(\vec{x})$. For the same
reason, we use the electric field $\ket{\vec{E}}$ and vacuum Maxwell
Green's function $\GG^{(0)}$, which are generalizations of the static
potential $\phi(\vec{x})$ and Coulomb kernel $v(\vec{x}, \vec{x}') =
1/(4\pi|\vec{x} - \vec{x}'|)$ common in quantum chemical treatments of
vdW interactions, as the former two include far-field EM retardation
effects. We again refer readers to~\appref{glossary} for more details.


Maxwell's equations~\eqref{Maxwellfield} may be formally inverted to
yield $\ket{\vec{E}} = \GG\ket{\vec{P}^{(0)}}$, where we define the
total Maxwell Green's function as the operator solving Maxwell's
equations in the presence of all susceptibilities:
\begin{equation} \label{eq:MaxwellGF}
  \left[\nabla\times\left(\nabla\times\right) -
  \frac{\omega^{2}}{c^{2}} (\II + \VV)\right]\GG =
  \frac{\omega^{2}}{c^{2}} \II.
\end{equation}
We point out that the assumption of reciprocal media implies that
$\GG$ is reciprocal, meaning $\GG = \GG^{\top}$ in position space,
i.e. $G_{ij}(\vec{x}, \vec{x}') = G_{ji}(\vec{x}',
\vec{x})$. Physically, this can be interpreted as leaving the physics
of an EM problem invariant if positions and polarizations of sources
and fields are interchanged. We further define the vacuum Maxwell
Green's function $\GG^{(0)}$ as the operator solving
Maxwell's equations in vacuum (i.e. $\VV = 0$): 
\begin{equation} \label{eq:MaxwellGvac}
  \left[\nabla\times\left(\nabla\times\right) -
  \frac{\omega^{2}}{c^{2}} \II\right]\GG^{(0)} =
  \frac{\omega^{2}}{c^{2}} \II.
\end{equation}

\section{General Scattering Among Molecular and Macroscopic
  Structures} \label{sec:generalsec}


In this section, we start with the most general formulation of EM
scattering among molecular and macroscopic bodies in order to derive
expressions for the vdW interactions and thermal radiation among
collections of such bodies, which we specifically do
in~\Secref{generalformulas}. These formulas depend only on the
T-operators describing the EM scattering properties and response of
individual bodies in isolation to all orders of scattering, and the
vacuum Green's function propagating fields between pairs of bodies in
a manner that only depends on the relative separations and
orientations of the bodies. While these formulas are not
new~\cite{RahiPRD2009, KrugerPRB2012}, they underscore the fact that
molecular and macroscopic bodies can be treated together, on the same
footing, in a unified formalism. In anticipation of our exposition of
the computational details of the description of molecular and
macroscopic DOFs, we then describe in~\Secref{DOFsplit} how to
equivalently rewrite the formulas for vdW interactions and thermal
radiation by partitioning the total response of the system into
molecular and macroscopic components. Finally, we give details about
the basis representations of molecular and macroscopic response
quantities in~\Secref{DOFmol} and~\Secref{DOFmac}, respectively, with
further derivations of the expression of $\GG^{(0)}$ in the molecular
basis in~\Secref{GvacGG}. We emphasize that although we focus in this
paper on vdW interactions and thermal radiation, the EM scattering
formalism is fully general, and the basis representation of molecular
response can be applied to problems including those involving
deterministic absorption or scattering, local density of
states~\cite{ReidIEEE2015, PolimeridisIEEE2015}, or fluorescent
emission~\cite{PolimeridisPRB2015}, among others.



After this section, the following three sections each deal with a
special case of the general formulas we present for vdW interactions
and thermal radiation. The first case is when the macroscopic bodies
do not change in separation or orientation relative to each other and
when consideration of heat transfer may be restricted just between
molecules. In the context of vdW interactions involving molecules,
there might be only one macroscopic body present, like a thick
metallic substrate or an atomic force microscopy (AFM) tip, in which case the question of
relative displacements or orientations among multiple macroscopic
bodies is moot. In the context of thermal radiation, consideration of
energy exchange may be restricted to molecules if again only one
macroscopic body is present, like an AFM tip, and it is in thermal
equilibrium with its environment, while the molecules are maintained
at a hotter temperature; this could be the case for measurements done
at room temperature on biological molecules in hotter samples of
living tissues or organisms. This special case allows for exploiting
the EM field response (Green's function) of the collection macroscopic
bodies without the molecules, which can be computed using a larger
variety of methods that do not make reference to T-operators. The
second case is an extension of the first case for extended molecular
structures that obey spatial periodic boundary conditions. In that
situation, we further derive analytical expressions for the vdW
interaction energy and thermal radiation among molecular bodies in the
presence of macroscopic bodies of commensurate periodicity, as well as
analytical formulas for the expression of $\GG^{(0)}$ in the set of
periodic molecular basis functions in a manner closely related to
Ewald summation. Such a situation could arise, for example, when
computing vdW interactions or thermal radiation for extended organic
molecular crystals like aspirin in the vicinity of planar or
periodically nanostructured metallic substrates. As we make clear in
those sections, however, our code only implements these classes of
computations for molecular bodies in vacuum or in the presence of a
single PEC plane for computational simplicity.
The third case, which is much more general, is for compact molecular
and macroscopic bodies when the macroscopic bodies are characterized
by spatially piecewise-constant permittivities, for which we may
reformulate our method to exploit the surface integral equation (SIE)
formulation of Maxwell's equations, as that yields significant
computational benefits in arbitrary macroscopic geometries over more
typical formulations, like spectral T-operator or volume integral
equation (VIE) formulations. Such a situation could arise, for
example, when considering vdW interactions or thermal radiation among
a collection of proteins, polynucleotides, compact low-dimensional
carbon allotropes, larger metallic nanoparticles, and an AFM tip, as
may be relevant in more complex novel biomedical settings. That said,
while we give the mathematical details of the method in this paper, we
have not yet implemented this functionality in our code and leave that
to future work.


\subsection{Scattering, vdW interactions, and thermal radiation
  among general polarizable bodies} \label{sec:generalformulas}


To start, we consider a collection of $N$ polarizable bodies labeled
$n \in \{1, \ldots, N\}$ with susceptibilities $\VV_{n}$. As we
clarify later, we assume that the electronic structures and
short-range interaction properties of each polarizable body are
unaffected by the presence of other bodies and that the bodies are
otherwise spatially disjoint, so the the total susceptibility $\VV =
\sum_{n = 1}^{N} \VV_{n}$ may be written as a direct sum over the
disjoint constituent susceptibilities, and $\VV_{n} = \PP_{n}
\VV\PP_{n}$ is written in terms of the projection operators $\PP_{n}$
onto the polarizable material DOFs of body $n$
; this means $\VV$ is block-diagonal in the space of polarizable 
bodies. With this in mind, Maxwell's equations may be written in
integral form as
\begin{equation} \label{eq:maxwellintegralform}
  \begin{split}
    \ket{\vec{E}} &= \ket{\vec{E}^{(0)}} + \GG^{(0)} \ket{\vec{P}} \\
    \ket{\vec{P}} &= \ket{\vec{P}^{(0)}} + \VV \ket{\vec{E}}
  \end{split}
\end{equation}
where $\ket{\vec{P}^{(0)}}$ are free polarization sources in the
polarizable 
bodies, while $\ket{\vec{E}^{(0)}}$ refers to incident fields produced
by sources outside of the system of polarizable bodies under
consideration (so it does not include the lowest-order radiated fields
$\GG^{(0)} \ket{\vec{P}^{(0)}}$, which are already accounted
in $\ket{\vec{P}}$). These equations can be self-consistently solved
to yield
\begin{equation} \label{eq:maxwellintegralsolve}
  \begin{split}
    \ket{\vec{P}} &= \TT \left(\VV^{-1} \ket{\vec{P}^{(0)}} +
    \ket{\vec{E}^{(0)}}\right) \\
    \ket{\vec{E}} &= \GG^{(0)} \TT \VV^{-1}
    \ket{\vec{P}^{(0)}} + \left(\II + \GG^{(0)}
      \TT\right)\ket{\vec{E}^{(0)}}
  \end{split}
\end{equation}
where we define the T-operator of the total system as $\TT^{-1} =
\VV^{-1} - \GG^{(0)}$, describing scattering to all orders within and
between all polarizable bodies; application of $\VV^{-1}$ to
$\ket{\vec{P}^{(0)}}$ is allowed as the susceptibilities are
nonsingular in the spaces spanned by the DOFs of the polarizable
bodies. We note that $\sum_{i,j} \partial_{i} \partial_{j}
T_{ij}(\vec{x}, \vec{x}')$ is exactly the fully interacting charge
density response in the nonretarded approximation (under the random
phase approximation), just as $\sum_{i,j} \partial_{i} \partial_{j}
V_{ij}(\vec{x}, \vec{x}')$ is the noninteracting charge density
response. While we emphasize that these polarization, scattering, and
radiation operators can be applied to a broad class of deterministic
as well as stochastic EM problems, in this paper we particularly
consider vdW interactions and thermal radiation.


Both vdW interactions and thermal radiation arise from quantum and
thermal fluctuations in the polarizations of material bodies. If the
free polarization sources $\ket{\vec{P^{(0)}}}$ and external incident
fields $\ket{\vec{E}^{(0)}}$ are taken to arise from quantum and
thermal fluctuations, then their correlations are given through the
fluctuation--dissipation theorem~\cite{Intravaia2011, Novotny2006}
(restoring explicit dependence on frequency for the sake of clarity)
\begin{equation} \label{eq:FDT}
  \begin{split}
    \bracket{\ket{\vec{P}^{(0)} (\omega)} \bra{\vec{P}^{(0)}
        (\omega')}} &= \frac{2\Theta(\omega, T)}{\omega}
    \asym(\VV(\omega)) \\ &\times 2\pi\delta(\omega - \omega') \\
    \bracket{\ket{\vec{E}^{(0)} (\omega)} \bra{\vec{E}^{(0)}
        (\omega')}} &= \frac{2\Theta(\omega, T)}{\omega}
    \asym(\GG^{(0)}(\omega)) \\ &\times 2\pi\delta(\omega -
    \omega')
  \end{split}
\end{equation}
which relates fluctuations in free polarizations or ambient vacuum
fields to dissipation quantities, respectively material absorption or
free-space far-field radiation; these are defined in terms of the
Planck factor $\Theta(\omega, T) =
(\hbar\omega/2)\coth\left(\hbar\omega/(2k_{\mathrm{B}}
T)\right)$~\footnote{The temperature $T$, in italics, is
  typographically distinct from the T-operator $\TT$, in blackboard
  font.}. We note also that the fluctuating free polarization sources
are uncorrelated from the ambient vacuum fields:
$\bracket{\ket{\vec{P}^{(0)}(\omega)}\bra{\vec{E}^{(0)}(\omega)}} =
\bracket{\ket{\vec{E}^{(0)}(\omega)}\bra{\vec{P}^{(0)}(\omega)}} =
0$. We point out that this is a generalization of the
fluctuation--dissipation theorem for the free charge density commonly
used in quantum chemical treatments of vdW interactions, where
$\bracket{\rho^{(0)}(\omega, \vec{x}) \rho^{(0)\star}(\omega,
  \vec{x}')} = \frac{2\Theta(\omega, T)}{\omega} \Im(\chi(\omega,
\vec{x}, \vec{x}')) \times 2\pi\delta(\omega - \omega')$ relates
fluctuations in the free charge density $\rho^{(0)}(\omega, \vec{x})$
to the dissipation given by the charge density response $\chi(\omega,
\vec{x}, \vec{x}')$~\cite{ZarembaPRB1976, EshuisTCA2012} (where it is
worth noting that this is to be distinguished from $\VV$, which is
often denoted as the susceptibility $\chi_{ij}(\omega, \vec{x},
\vec{x}')$ in continuum EM literature).


The total vdW free energy in a system of polarizable bodies at thermal
equilibrium may be written in a Hellmann--Feynman form as the
interaction between the total polarizations and
fields~\cite{AgarwalPRA1975, HermannCR2017}:
\begin{equation}
  \mathcal{F}_{\mathrm{tot}} = -\int_{0}^{1} \int \bracket{\vec{P}(t,
    \vec{x}) \cdot \vec{E}(t, \vec{x})}~\mathrm{d}^{3}
  x~\frac{\mathrm{d}\lambda}{\lambda}
\end{equation}
where $\lambda$ is the Hellmann--Feynman adiabatic connection
parameter which linearly rescales $\VV$ and $\GG^{(0)}$, and
where the expectation value $\bracket{\ldots}$ is taken over time, or
equivalently over ensembles by ergodicity. By writing the
polarizations and fields in the frequency domain $\vec{P}(t, \vec{x})
= \int_{-\infty}^{\infty} \vec{P}(\omega, \vec{x}) e^{-\im\omega
  t}~\frac{\mathrm{d}\omega}{2\pi}$ and $\vec{E}(t, \vec{x}) =
\int_{-\infty}^{\infty} \vec{E}(\omega, \vec{x}) e^{-\im\omega
  t}~\frac{\mathrm{d}\omega}{2\pi}$ (where in a slight abuse of
notation, the same symbol is used for time and frequency domain
quantities), we may then rewrite the total vdW energy as
\begin{multline}
  \mathcal{F}_{\mathrm{tot}} = -\int_{0}^{1} \int_{-\infty}^{\infty}
  \int_{-\infty}^{\infty} \bracket{\trace[\ket{\vec{E}(\omega)}
      \bra{\vec{P}(\omega')}] e^{-\im(\omega -
      \omega')t}} \times \\ \frac{\mathrm{d} \omega~\mathrm{d}
    \omega'}{(2\pi)^{2}}~\frac{\mathrm{d} \lambda}{\lambda}
\end{multline}
in a basis-independent manner. Using the results
of~\eqref{maxwellintegralsolve} (where rescaling by $\lambda$ is
implicit for now), algebraic manipulations yield
$\bracket{\trace[\ket{\vec{E}(\omega)} \bra{\vec{P}(\omega')}]
  e^{-\im(\omega - \omega')t}} = \frac{2\Theta(\omega, T)}{\omega}
\trace[\asym(\GG^{(0)}(\omega) \TT(\omega))] \times 2\pi\delta(\omega
- \omega')$; plugging this into the formula for
$\mathcal{F}_{\mathrm{tot}}$ allows for reduction of the integration
to a single frequency variable instead of two, so henceforth frequency
dependence will again be implicit in the notation. Additionally,
restoring the factors of the adiabatic coupling coefficient $\lambda$
means $\asym(\GG^{(0)} \TT) = \asym\left(\sum_{n = 1}^{\infty}
\lambda^{2n} (\GG^{(0)} \VV)^{n}\right)$, so the integration over
$\lambda$ can be done as $\int_{0}^{1} \lambda^{2n -
  1}~\mathrm{d}\lambda = \frac{1}{2n}$, while $\tr[\asym(\mathbb{A})]
= \Im(\tr[\mathbb{A}])$ for any operator $\mathbb{A}$ means the
imaginary part operation can be applied to the whole integral; it can
then be seen that $-\sum_{n = 1}^{\infty} \frac{1}{n} (\GG^{(0)}
\VV)^{n} = \tr[\ln[\II - \GG^{(0)} \VV]] = \ln(\det[\II - \GG^{(0)}
  \VV])$. Moreover, causality means that $\VV(-\omega^{\star}) =
\VV^{\star}(\omega)$ and $\GG^{(0)}(-\omega^{\star}) =
\GG^{(0)\star}(\omega)$, so $\int_{-\infty}^{\infty}
f(\omega)~\mathrm{d}\omega = 2~\int_{0}^{\infty}
f(\omega)~\mathrm{d}\omega$. This therefore allows for writing the
total vdW free energy as
\begin{equation}
  \mathcal{F}_{\mathrm{tot}} = \frac{1}{\pi}
  \Im\left(\int_{0}^{\infty} \frac{\Theta(\omega, T)}{\omega}
  \ln(\det[\II - \GG^{(0)} \VV])~\mathrm{d}\omega\right)
\end{equation}
where all quantities in the log-determinant expression depend on
$\omega$; we note that this formula reduces to the adiabatic
connection formula for the vdW interaction energy in the nonretarded
regime, performing an integration by parts using the definitions of
the charge density response and $\GG^{(0)}$ in terms of $\VV$ and the
Coulomb kernel, respectively. Hence, there are two final steps needed
to reach the desired expressions for the vdW interaction free
energy. The first is that the interaction energy is the difference
between the total energy in two different geometric configurations. If
we maintain the assumption that the material properties of each
polarizable body do not change with respect to geometric
configuration, so that the susceptibilities $\VV_{n}$ only change
trivially by virtue of rigid geometric transformations, then we can
identify $\II - \GG^{(0)} \VV = \TT^{-1} \VV$, define $\TT_{\infty}$
as the T-operator corresponding to each object in isolation (which by
assumption only affects long-range EM scattering among the various
bodies), and then rewrite the differences in the integrands between
the desired configuration and the reference configuration of each
object in isolation as $\ln(\det[\TT_{\infty} \TT^{-1}])$. The second
is that while the above frequency integral may be evaluated directly,
it is analytically and numerically more desirable to perform a Wick
rotation to positive imaginary frequency, where $\Theta(\omega, T)$
has simple poles whose residues may be evaluated
easily~\cite{RodriguezPRA2007, Intravaia2011}. Thus, we derive the vdW
interaction free energy as
\begin{equation} \label{eq:vdWfreeenergy}
  \mathcal{F} = k_{\mathrm{B}} T \sum_{l = 0}^{\infty} {}'
  \ln(\det[\TT_{\infty} \TT^{-1}])
\end{equation}
where all quantities are evaluated at the Matsubara frequencies
$\omega_{l} = \im\xi_{l}$ for $\xi_{l} = \frac{2\pi k_{\mathrm{B}}
  Tl}{\hbar}$, and where the prime indicates a weight of $1/2$ at $l =
0$ to avoid double-counting (as this was originally an integral over
the entire real frequency axis). We point out that this Matsubara
summation procedure is mathematically like a Riemann sum, and as $T
\to 0$, this sum converges to an integral. Indeed, as $T \to 0$, the
entropic contributions to the free energy vanish, and we recover the
familiar expression for the zero-temperature vdW interaction energy
\begin{equation}
  \mathcal{E} = \hbar \int_{0}^{\infty} \ln(\det[\TT_{\infty}(\im\xi)
    \TT^{-1}(\im\xi)])~\frac{\mathrm{d}\xi}{2\pi}
\end{equation}
though through the rest of this paper, we will use the notation
$\mathcal{F}(T)$ to denote the vdW interaction free energy of a given
system at temperature $T$ (so interactions at $T = 0$ will be denoted
$\mathcal{F}(0)$).



We now turn to thermal emission and RHT among
polarizable bodies. Each of these phenomena can be described as the
net work (with respect to relevant temperature differences) done by
fields on the polarizations of one body labeled $n$, where those
fields have been radiated by fluctuating sources in another body
labeled $m$ (which may be the same as $n$). Consequently, the ambient
fluctuating vacuum fields $\ket{\vec{E}^{(0)}}$
in~\eqref{maxwellintegralsolve} are irrelevant and may therefore be
neglected, while the fluctuating sources are $\ket{\vec{P}^{(0)}_{m}}
= \PP_{m} \ket{\vec{P}^{(0)}}$. To start, the total power may be
written as
\begin{equation}
  P = \int \bracket{\vec{J}(t, \vec{x}) \cdot \vec{E}(t,
    \vec{x})}~\mathrm{d}^{3} x
\end{equation}
via Poynting's theorem, where $\vec{J}(t, \vec{x}) =
\frac{\partial}{\partial t} \vec{P}(t, \vec{x})$, and where the
expectation value $\bracket{\ldots}$ may again be considered over time
or equivalently over ensembles through ergodicity. Using the prior
expressions for the Fourier transforms as well as the projection
operators allows for rewriting
\begin{equation}
  P = \int_{-\infty}^{\infty} \int_{-\infty}^{\infty}
  \bracket{\trace[\PP_{n} \ket{\vec{E}(\omega)} \bra{\vec{J}(\omega')}
      \PP_{n}]e^{-\im(\omega -
      \omega')t}}~\frac{\mathrm{d}\omega~\mathrm{d}\omega'}{(2\pi)^{2}}
\end{equation}
in a basis-independent manner. Following similar steps as with the vdW
derivation, using the fact that $\VV$ is block-diagonal and is
therefore invertible in the space of material DOFs, algebraic
manipulations yield $\bracket{\trace[\PP_{n} \ket{\vec{E}(\omega)}
    \bra{\vec{J}(\omega')} \PP_{n}]e^{-\im(\omega - \omega')t}} =
2\im\Theta(\omega, T_{m}) \PP_{n} \GG^{(0)} \TT \PP_{m}
\asym(\VV_{m}^{-1\dagger}) \PP_{m} \TT^{\dagger} \PP_{n} \times
2\pi\delta(\omega - \omega')$, where all quantities depend on
$\omega$. Plugging this in and again using the causality properties of
the relevant response quantities to reduce the integral over
frequencies to the positive axis allows for writing the power as
\begin{multline}
  P = -\int_{-\infty}^{\infty} 4~\Theta(\omega, T_{m}) \times
  \\ \trace[\asym(\VV_{m}^{-1\dagger}) \PP_{m} \TT^{\dagger}
    \asym(\PP_{n} \GG^{(0)})
    \TT\PP_{m}]~\frac{\mathrm{d}\omega}{2\pi}.
\end{multline}
We note that this only depends on the temperature of one of the bodies
in question, and not that of another body or the ambient
environment. In general, we can still define the dimensionless
radiation spectrum from body $m$ to body $n$ at each frequency as
\begin{equation} \label{eq:generalenergytransfer}
  \Phi^{(m)}_{n} = -4~\trace[\asym(\VV_{m}^{-1\dagger})
    \PP_{m} \TT^{\dagger} \asym(\PP_{n} \GG^{(0)})
    \TT\PP_{m}]
\end{equation}
and then, in terms of that, define general frequency integrated power
quantities $P = \int_{0}^{\infty}
W(\omega)~\frac{\mathrm{d}\omega}{2\pi}$, where $W$ is defined as
\begin{equation}
  W^{(m)} = \sum_{n = 1}^{N} s_{nm} \Phi^{(m)}_{n} (\Theta(\omega,
  T_{n}) - \Theta(\omega, T_{\mathrm{env}}))
\end{equation}
for thermal emission of body $m$ into an environment of ambient
temperature $T_{\mathrm{env}}$ in terms of the sign function $s_{nm} =
1 - 2\delta_{nm}$, or as
\begin{equation}
  W_{m\to n} = \Phi^{(m)}_{n} (\Theta(\omega, T_{m}) - \Theta(\omega,
  T_{n}))
\end{equation}
for RHT between bodies $m$ and $n$. In the context
of thermal emission and RHT, as the Planck
function $\Theta$ only appears in the form of differences at different
temperatures, the zero-point contribution $\hbar\omega/2$ drops out,
so it is helpful to redefine $\Theta(\omega, T) =
\hbar\omega/(\operatorname{exp}(\hbar\omega/(k_{\mathrm{B}} T)) - 1)$
without the zero-point term.

\subsection{Partitioning molecular and macroscopic DOFs} \label{sec:DOFsplit}



None of the formulas in the prior subsection made particular reference
to whether the polarizable bodies were atomistic or continuous, nor to
any particular basis set, but as will become clear shortly, it is
useful for the purposes of physical interpretation and computational
convenience to introduce that distinction. That said, before
proceeding, we clarify that the terms ``molecular'' and
``macroscopic'' are not absolute descriptors, but depend on the
details of the configuration of polarizable bodies. As a general rule
of thumb, bodies that are smaller than about $5~\mathrm{nm}$ in at
least one dimension or in feature size must be treated in an ab-initio
manner incorporating atom-scale effects. We call such bodies
``molecular'' and use the label ``mol'' as a superscript or subscript
associated with relevant response quantities, because prior work has
focused on compact molecules and finite-size low-dimensional atomistic
systems. As even certain structures of infinite extent in multiple
dimensions must generally be treated atomistically, we refer to such
structures as ``molecular'' for semantic consistency; as an example,
graphene can arguably be visualized as a polycyclic aromatic
hydrocarbon of infinite extent with no termination points where
hydrogen atoms may lie. If none of the above conditions hold, then
bodies may be well-described by coarse-grained continuum local bulk
models of material susceptibility; even at such small length scales
(e.g. metallic spherical nanoparticles), we term such bodies
``macroscopic'' due to the accuracy of bulk material modeling, and
associate the label ``mac'' as a superscript or subscript associated
with relevant response quantities. Even this categorization is not
complete, because as the distances between proximate bodies fall below
$O(1~\mathrm{nm})$, each body would need to be treated atomistically;
this would apply even to bulk metal substrates, at least with respect
to the atoms closest to the other body. Thus, we assume the validity
of continuum models for macroscopic bodies that are at least
$5~\mathrm{nm}$ along each dimension and feature and at least
$1~\mathrm{nm}$ away from any other body; if these conditions are
violated, the ``macroscopic'' body would need to be treated
atomistically, but this can anyway be done in our formalism.


Formally, we separate $\VV = \VV_{\mathrm{mol}} + \VV_{\mathrm{mac}}$
as a sum of susceptibilities for disjoint collections of objects,
where $\VV_{\mathrm{mol}} = \sum_{s = 1}^{N_{\mathrm{mol}}} \VV_{s}$
and $\VV_{\mathrm{mac}} = \sum_{a = 1}^{N_{\mathrm{mac}}} \VV_{a}$ are
each block-diagonal in their respective sets of DOFs. As described
above, the assumption of disjointness will hold for any pair of bodies
that are sufficiently separated and of appropriate dimensionality that
short-range electronic exchange and correlation effects may be
neglected for material DOFs between the two bodies. With this, we
further define block $2\times 2$ matrices via this separation of
molecular (top block row and left block column) from macroscopic
(bottom block row and right block column) DOFs:
\begin{equation} \label{eq:blockVandT}
  \begin{split}
    \VV &= \begin{bmatrix}
      \VV_{\mathrm{mol}} & 0 \\
      0 & \VV_{\mathrm{mac}}
    \end{bmatrix} \\
    \TT^{-1} &= \begin{bmatrix}
      \TT_{\mathrm{mol}}^{-1} & -\GG^{(0)} \\
      -\GG^{(0)} & \TT_{\mathrm{mac}}^{-1}
    \end{bmatrix}
  \end{split}
\end{equation}
where in the blocks of $\TT^{-1}$, $\TT_{\mathrm{mol(mac)}}^{-1} =
\VV_{\mathrm{mol(mac)}}^{-1} - \GG^{(0)}$ encodes scattering
properties of the collection of molecules (macroscopic bodies) in a
particular geometric configuration relative to each other in vacuum in
the absence of macroscopic bodies (molecules), while the off-diagonal
blocks $\GG^{(0)}$ propagate EM fields in vacuum between molecular and
macroscopic DOFs. These formulas can simplify physical interpretation
and computational implementation of various deterministic as well as
stochastic EM phenomena involving molecules in conjunction with
macroscopic bodies, though we specifically focus on vdW interactions
as well as thermal emission and thermal radiation.

In general, the vdW interaction free energy or force may be desired in
situations where one or more molecular or macroscopic bodies are taken
together as a compound object; for example, if the force on an AFM tip
in proximity with a graphene sheet adsorbed at a particular small
separation to a metallic surface is desired, then the reference
configuration would be the tip isolated from the graphene and metal
surfaces, but the graphene sheet would remain at the same small
separation from the adsorbent metallic surface. In such a case, the
relevant reference configuration would not correspond to every body
being isolated from each other in vacuum. In analogy
to~\eqref{blockVandT}, we may define the block matrix $\TT_{\infty}$
as the reference configuration of molecular and macroscopic bodies via
\begin{equation}
  \TT_{\infty}^{-1} = \begin{bmatrix}
    \TT_{\mathrm{mol}\infty}^{-1} & -\GG^{(0)} \\
    -\GG^{(0)} & \TT_{\mathrm{mac}\infty}^{-1}
  \end{bmatrix}
\end{equation}
where $\TT_{\mathrm{mol(mac)}\infty}$ encode the positions,
displacements, and orientations among only molecular or macroscopic
bodies in the given reference configuration; for example, if the
reference configuration is of all bodies infinitely separated, the
representation of the off-diagonal blocks of $\GG^{(0)}$ will
vanish. Expanding $\TT_{\infty}$ and the determinant in the vdW
summand blockwise leads to an expression for the summand
\begin{equation} \label{eq:vdWintegrand}
  \begin{split}
    \ln(\det(\TT_{\infty} \TT^{-1})) &=
    \ln(\det(\TT_{\mathrm{mac}\infty} \TT_{\mathrm{mac}}^{-1})) \\ &+
    \ln\left(\frac{\det(\TT_{\mathrm{mol}}^{-1} - \GG^{(0)}
      \TT_{\mathrm{mac}}
      \GG^{(0)})}{\det(\TT_{\mathrm{mol}\infty}^{-1} - \GG^{(0)}
      \TT_{\mathrm{mac}\infty} \GG^{(0)})}\right)
  \end{split}
\end{equation}
where conceptually, the first term is the vdW interaction energy
purely among macroscopic bodies in vacuum (in the absence of
molecules) relative to their reference configuration, while the second
term is the vdW interaction energy among molecular bodies in a
scattering background created by the macroscopic bodies, relative to
their reference configuration accounting for the change in the
macroscopic bodies' positions and orientations from the corresponding
reference configuration too. This formula has the additional benefit
of making explicit the full interchangeability of molecular and
macroscopic DOFs, as a fully mathematically equivalent formula arises
simply by exchanging the labels
$\mathrm{mol}~\leftrightarrow~\mathrm{mac}$ and associated basis
functions, showing how our formulation really does treat molecular and
macroscopic DOFs on an equal footing; in particular, performing this
exchange allows for writing the vdW interaction energy as the sum of
that purely between molecular bodies in vacuum (in the absence of
macroscopic bodies) relative to their reference configuration and the
vdW interaction energy among macroscopic bodies in a scattering
background created by the molecular bodies, relative to their
reference configuration accounting for the change in the molecular
bodies' positions and orientations too.

The formula for thermal emission and RHT
in~\eqref{generalenergytransfer} holds for general molecular or
macroscopic bodies, treating both sorts of bodies on the same footing
and featuring the same benefits and pitfalls
as~\eqref{vdWfreeenergy}. For the same reason, it may be more
beneficial to explicitly separate molecular from macroscopic DOFs as
in~\eqref{blockVandT}. In fact, the possibility of energy exchange
between molecules and macroscopic bodies, going beyond energy exchange
between molecules or macroscopic bodies alone, allows for such a
separation to more clearly illustrate the richness of the mathematical
formalism and computational \& physical implications. Physically, heat
transfer among macroscopic bodies in the presence of molecules can be
realized via molecular junctions, heat transfer between a macroscopic
body and a molecule could be realized via a metallic probe or
nanoparticle locally heating a cancerous protein, and heat transfer
between molecular bodies in the presence of macroscopic bodies could
be seen in energy exchange between a hot graphene sheet and a cooler
fullerene in the vicinity of a thick metallic substrate. Therefore, it
behooves us to more fully draw out the formulas for thermal radiation
in each such case. The radiation spectrum between just molecules $m$
and $n$, in the presence of other molecules and macroscopic bodies,
may be written as
\begin{equation} \label{eq:heattransferonlymolormac}
  \begin{split}
    \Phi^{(m)}_{n} = -4~\trace\Big[&\asym(\VV_{m}^{-1\dagger}) \times
      \\ &\PP_{m} (\TT_{\mathrm{mol}}^{-1\dagger} - \GG^{(0)\dagger}
      \TT_{\mathrm{mac}}^{\dagger} \GG^{(0)\dagger})^{-1} \times
      \\ &\asym(\PP_{n} (\GG^{(0)} + \GG^{(0)} \TT_{\mathrm{mac}}
      \GG^{(0)})) \times \\ &(\TT_{\mathrm{mol}}^{-1} - \GG^{(0)}
      \TT_{\mathrm{mac}} \GG^{(0)})^{-1} \PP_{m}\Big]
  \end{split}
\end{equation}
after using~\eqref{blockVandT} and performing further operator
manipulations, showing that the macroscopic bodies merely form a
scattering background for energy exchange among molecular
bodies. Likewise, the radiation spectrum between just macroscopic
bodies $m$ and $n$ may be written exactly
as~\eqref{heattransferonlymolormac} after exchanging the labels
$\mathrm{mol}~\leftrightarrow~\mathrm{mac}$, showing again that the
molecules merely form a scattering background for energy exchange
among the macroscopic bodies. Finally, if $m$ is a molecular body
while $n$ is a macroscopic body, the heat transfer may be written
(with the reverse again obtained under the substitution $\mathrm{mol}
\leftrightarrow \mathrm{mac}$) as:
\begin{equation} \label{eq:heattransferbothmolandmac}
  \begin{split}
    \Phi^{(m)}_{n} = 4~\trace\Big[&\asym(\VV_{m}^{-1\dagger}) \times \\ &\PP_{m}
      \TT_{\mathrm{mol}}^{\dagger} \GG^{(0)\dagger}
      (\TT_{\mathrm{mac}}^{-1\dagger} - \GG^{(0)\dagger}
      \TT_{\mathrm{mol}}^{\dagger} \GG^{(0)\dagger})^{-1} \times \\ 
      &\PP_{n} \asym(\VV_{n}^{-1\dagger}) \PP_{n} \times \\
      &(\TT_{\mathrm{mac}}^{-1} - \GG^{(0)} \TT_{\mathrm{mol}}
      \GG^{(0)})^{-1} \GG^{(0)} \TT_{\mathrm{mol}} \PP_{m}\Big]
  \end{split}
\end{equation}
after manipulating operators and using the fact that the
susceptibility operators are block-diagonal, so $\PP_{n} \VV = \PP_{n}
\VV_{n} \PP_{n}$; this expression clearly shows symmetry in the
equation when the molecular and macroscopic bodies are
interchanged. Note that while~\eqref{FDT} and past T-operator and VIE
formulations of thermal radiation make use of $\VV$, we choose to
write our expressions in terms of $\VV^{-1}$ as much as possible,
because as we will shortly make clear, the molecular basis expansion
we use directly gives $\VV_{\mathrm{mol}}^{-1}$ without need for
further inversion.


The only thing remaining to describe scattering among
molecular and macroscopic bodies is to represent
$\TT_{\mathrm{mol(mac)}}$ in appropriate basis sets and the
off-diagonal blocks $\GG^{(0)}$ in the basis functions connecting
molecular and macroscopic bodies. Such a representation will make the
practical computational aspects and physical interpretations of
formulas for vdW interactions and thermal radiation more clear.

\subsection{Basis expansions of molecular DOFs} \label{sec:DOFmol}




We write the molecular susceptibility as $\VV_{\mathrm{mol}} =
\sum_{pi,qj} \alpha_{pi,qj} \ket{\vec{f}_{pi}}\bra{\vec{f}_{qj}}$. In
general, the molecular susceptibility must account for the
contributions of electrons, phonons, and other collective modes to the
response; especially in metallic systems, this typically requires
delocalized basis functions $\ket{\vec{f}_{pi}}$. However, for
insulating or weakly conducting molecular systems, we may model the
molecule as being made of nuclei that are harmonically coupled to
nearest neighbors within each molecule and effective valence
electronic harmonic oscillators associated 1-to-1 with a corresponding
nucleus; this accurately captures the features of molecular response
salient to fluctuational EM phenomena, like vdW interactions and
thermal radiation, at ultraviolet frequencies via the effective
electronic oscillators and at infrared frequencies via phonons arising
from the coupled nuclear oscillators, and is valid for low
temperatures where the harmonic approximation holds. In particular,
the molecules together have $N$ atoms labeled $p$ located at positions
$\vec{r}_{p}$, each of which has an effective electronic oscillator of
charge $q_{\mathrm{e}p}$ and mass $m_{\mathrm{e}p}$ (which might not
be equal to the fundamental electron charge or mass), coupled to its
corresponding nucleus via an isotropic harmonic spring of constant
$k_{\mathrm{e}p}$ and damped isotropically with coefficient
$b_{\mathrm{e}p} = m_{\mathrm{e}p} \gamma_{\mathrm{e}p}$ (written in
terms of a damping \emph{rate} $\gamma_{\mathrm{e}p}$), and a nucleus
of mass $m_{\mathrm{I}p}$ coupled (in addition to its own electronic
oscillator) to its nearest neighbors within each molecule via
anisotropic spring constants $\mathbb{K}_{pq}$ and damped
isotropically with coefficient $b_{\mathrm{I}p} = m_{\mathrm{I}p}
\gamma_{\mathrm{I}p}$. The quantities $q_{\mathrm{e}p}$,
$m_{\mathrm{e}p}$, $k_{\mathrm{e}p}$, $\mathbb{K}_{pq}$, and atomic
coordinates $\vec{r}_{p}$ are computed for each molecular body (or
cluster, if a set of molecular bodies exhibits a more strongly
correlated electronic structure even for nuclear separations beyond a
few bond lengths) separately via density functional theory (DFT)
calculations in conjunction with Hirshfeld
partitioning~\cite{HermannCR2017, TkatchenkoJCP2013, DiStasioJPCM2014,
  TkatchenkoADFM2014}
, while $m_{\mathrm{I}p}$ is given from elemental data, and the
damping rates $\gamma_{\mathrm{e}p}$ and $\gamma_{\mathrm{I}p}$ are
taken from empirical data. As the molecular DOFs are all damped
coupled harmonic oscillators, with only the effective electronic
oscillators directly coupling to electric fields (neglecting the
nonlinear magnetic contribution to the Lorentz force, as may be done
at typical operating temperatures as the relevant speeds of the
material DOFs are nonrelativistic), the frequency domain equations of
motion are simply
\begin{equation} \label{eq:molEqOM}
  \begin{split}
    & \begin{bmatrix}
      K_{\mathrm{e}} - \im\omega B_{\mathrm{e}} - \omega^{2} M_{\mathrm{e}} & -K_{\mathrm{e}} \\
      -K_{\mathrm{e}} & K_{\mathrm{e}} + K_{\mathrm{I}} - \im\omega B_{\mathrm{I}} - \omega^{2} M_{\mathrm{I}}
    \end{bmatrix}
    \begin{bmatrix}
      x_{\mathrm{e}} \\
      x_{\mathrm{I}}
    \end{bmatrix}
    \\ &=
    \begin{bmatrix}
      Q_{\mathrm{e}} e_{\mathrm{e}} \\
      0
    \end{bmatrix}
  \end{split}
\end{equation}
where
$(Q_{\mathrm{e}}, M_{\mathrm{e}}, M_{\mathrm{I}}, B_{\mathrm{e}},
B_{\mathrm{I}}, K_{\mathrm{e}}, K_{\mathrm{I}})$ collect the
parameters $q_{\mathrm{e}p}$, $m_{\mathrm{e}p}$, $m_{\mathrm{I}p}$,
$b_{\mathrm{e}p}$, $b_{\mathrm{I}p}$, $k_{\mathrm{e}p}$, and
$\mathbb{K}_{pq}$ respectively into $3N\times 3N$ matrices. These
equations of motion determine the nuclear displacements
$x_{\mathrm{I}}$ and electronic dipole moments
$p_{\mathrm{e}} = Q_{\mathrm{e}} x_{\mathrm{e}}$ in response to an
electric field $e_{\mathrm{e}}$ obtained by evaluating $\ket{\vec{E}}$
at the atomic positions $\vec{r}_{p}$ (leading to a $3N$-dimensional
vector); note that in this model, only the electronic oscillators
directly couple to the electric field. Solving for
$p_{\mathrm{e}} = \alpha e_{\mathrm{e}}$ gives the susceptibility
matrix
\begin{equation} \label{eq:molalphamat}
  \begin{split}
    \alpha &= Q_{\mathrm{e}} \Bigg(K_{\mathrm{e}} - \im\omega B_{\mathrm{e}} - \omega^{2} M_{\mathrm{e}} \\ &- K_{\mathrm{e}} \left(K_{\mathrm{e}} + K_{\mathrm{I}} - \im\omega B_{\mathrm{I}} - \omega^{2} M_{\mathrm{I}}\right)^{-1} K_{\mathrm{e}}\Bigg)^{-1} Q_{\mathrm{e}}
  \end{split}
\end{equation}
entering the basis expansion of $\VV_{\mathrm{mol}}$. The
distinction between the ultraviolet contributions primarily from the
electronic oscillators and the infrared contributions primarily from
phonons arises due to $M_{\mathrm{e}}$ and $M_{\mathrm{I}}$ differing
by 4 orders of magnitude, in contrast to the comparable magnitude of
$K_{\mathrm{e}}$ to $K_{\mathrm{I}}$. Additionally, as we model the
electrons and nuclei as harmonic oscillators, we use Gaussian basis
functions
\begin{equation} \label{eq:molGaussfunc}
  \vec{f}_{pi} (\vec{x}) = \left(\sqrt{2\pi} \sigma_{p}\right)^{-3} \exp\left(-\frac{(\vec{x} - \vec{r}_{p})^{2}}{2\sigma_{p}^{2}}\right) \vec{e}_{i}
\end{equation}
where the widths $\sigma_{p}$, rather than being phenomenological,
microscopically capture the nonlocal response of each molecule at each
frequency by virtue of the definition $\sigma_{p} (\omega) =
(\alpha_{p}(\omega) / 3)^{1/3} / (2\sqrt{\pi})$~\cite{MayerPRB2007,
  VenkataramPRL2017, VenkataramPRL2018, VenkataramSCIADV2019} in terms
of $\alpha_{p} (\omega) = |\sum_{q,j} \alpha_{pj,qj}
(\omega)|/3$. This choice of contracting the molecular susceptibility
and averaging over the Cartesian tensor components to yield isotropic
atomic fragment polarizabilities is consistent with previous
expressions for isotropic local molecular susceptibilities used to
construct Gaussian basis functions in the absence of
phonons~\cite{HermannCR2017, DiStasioJPCM2014, AmbrosettiJCP2014,
  VenkataramPRL2017}, and is also consistent with similar expressions
deriving atomic polarizabilities from screened molecular
susceptibilities~\cite{HermannCR2017, DiStasioJPCM2014,
  AmbrosettiJCP2014} (T-operators, though those by definition include
long-range EM interactions, unlike our bare expressions for
$\VV$). Physically, this definition accounts not only for the change
in the response at any given atom due to nonlocal internuclear
couplings $K_{\mathrm{I}}$, but also for the full spatial extent of
the nonlocality by summing over contributions from other atoms as
well, though it does not explicitly preserve the anisotropy of the
response in the Gaussian widths; the latter point, which could become
especially important for low-dimensional materials like carbyne or
graphene, is not further addressed in this work, but will be the
subject of future work. Mathematically, we have found that while at
imaginary frequency (relevant to vdW interactions) the
polarizabilities $\alpha_{p} (\im\xi)$ will always be positive, at
real frequency (relevant to thermal radiation and other EM scattering
phenomena), the absolute value is necessary to ensure real positive
Gaussian widths when constructing the basis functions, because the
polarizability matrix $\alpha$ will in general be complex-valued and
will have some diagonal or off-diagonal elements that have negative
real parts at frequencies above electronic or phononic
resonances. Additionally, the summation over other atoms $q$ (as
opposed to an alternative like $|\sum_{j} \alpha_{pj,pj} (\omega)|/3$
which only accounts for the response at a given atom) is necessary to
ensure positive-definiteness of $\asym(\TT)$ at real $\omega$ (or of
$\TT$ at $\omega = \im\xi$) in the molecular basis, though we have not
been able to conclusively prove this statement. We point out that for
most compact molecules as well as extended low-dimensional structures,
the Gaussian widths will be less than $O(5~\mathrm{nm})$.

We also require computation of the matrix elements
$\bracket{\vec{f}_{pi}|\GG^{(0)} \vec{f}_{qj}}$ for atoms within and
between molecules, in order to represent $\TT_{\mathrm{mol}}$. The use
of Gaussian basis functions fortunately leads to analytical
expressions for these matrix elements, which we first state and
qualitatively discuss here, deriving these expressions shortly
afterwards. The expression
\begin{equation} \label{eq:molGvacGG}
  \begin{split}
    &\bracket{\vec{f}_{pi}|\GG^{(0)} \vec{f}_{qj}} =
    (\partial_{r_{pi}} \partial_{r_{pj}} + (\omega/c)^{2} \delta_{ij})
    \times \\ &\frac{\exp\left(-q^{2} / 4\right)}{8\pi
      |\vec{r}_{p} - \vec{r}_{q}|} \Bigg[e^{\im\rho q}
      \erfc\left(-\frac{\im q}{2} - \rho\right) - e^{-\im\rho q}
      \erfc\left(-\frac{\im q}{2} + \rho\right)\Bigg]
  \end{split}
\end{equation}
is written analytically in terms of the dimensionless quantities $q
\equiv (\omega/c)\sqrt{2(\sigma_{p}^{2} + \sigma_{q}^{2})}$ and $\rho
\equiv |\vec{r}_{p} - \vec{r}_{q}|/\sqrt{2(\sigma_{p}^{2} +
  \sigma_{q}^{2})}$, thereby obviating the need for time-consuming
numerical cubature over the volumes of the basis functions and in turn
speeding up evaluation of the basis representation of
$\TT_{\mathrm{mol}}$. We note that while our choice of basis functions
$\ket{\vec{f}_{pi}}$ is effectively a Galerkin discretization
reminiscent of VIE formulations of Maxwell's equations, the number of
basis functions is determined directly by the number of atoms as
opposed to being chosen arbitrarily for numerical convergence. For
nonzero Gaussian widths $\sigma_{p}^{2} + \sigma_{q}^{2}$, these
matrix elements are finite even in the coincidence limit $|\vec{r}_{p}
- \vec{r}_{q}| \to 0$, unlike those of $\GG^{(0)} (\omega, \vec{x},
\vec{x}')$, though the latter can be attained in the limit
$\sigma_{p}^{2} + \sigma_{q}^{2} \to 0$; this approach to a finite
value captures the screening of long-range EM interactions due to
short-range electronic response.

\subsection{Basis expansions of macroscopic DOFs} \label{sec:DOFmac}
For macroscopic bodies treated using continuous dielectric functions,
while the expression $\TT_{\mathrm{mac}}^{-1} =
\VV_{\mathrm{mac}}^{-1} - \GG^{(0)}$ can technically be used, it is
not necessarily the most efficient way to obtain a basis
representation $\{\ket{\vec{b}_{\beta}}\}$ for
$\TT_{\mathrm{mac}}$. In particular, while VIE methods do use this
expression in conjunction with localized voxel or
Schaubert--Wilton--Glisson basis functions $\{\ket{\vec{b}_{\beta}}\}$
to represent $\TT_{\mathrm{mac}}$, other methods like scattering
methods in planar or spherical waves, or finite-difference methods,
may represent $\TT_{\mathrm{mac}}$ in a way that is mathematically
equivalent, but less obviously so, to the above expression. In any
case, the choice of macroscopic basis $\ket{\vec{b}_{\beta}}$ will
also affect the computation and convergence properties of the
representation of $\GG^{(0)}$ as $\bracket{\vec{b}_{\beta}|\GG^{(0)}
  \vec{f}_{qj}}$ connecting molecular and macroscopic DOFs; the term
$\GG^{(0)} \ket{\vec{f}_{qj}}$ is evaluated analytically in position
space in the same way as $\bracket{\vec{f}_{pi}|\GG^{(0)}
  \vec{f}_{qj}}$ but in the limit $\sigma_{p} \to 0$ and with
$\vec{r}_{p}$ replaced by a generic $\vec{x}$, as those limits applied
to $\ket{\vec{f}_{pi}}$ yield a Dirac delta function, while the
convergent, smooth, analytic properties of $\GG^{(0)}
\ket{\vec{f}_{qj}}$ facilitate analytical or numerical evaluation of
the matrix elements $\bracket{\vec{b}_{\beta}|\GG^{(0)}
  \vec{f}_{qj}}$. Thus, the choice of macroscopic basis
$\ket{\vec{b}_{\beta}}$ should in practice account for the convergence
properties of both $\TT_{\mathrm{mac}}$ and
$\bracket{\vec{b}_{\beta}|\GG^{(0)} \vec{f}_{qj}}$.

\subsection{Expression of $\GG^{(0)}$ in the molecular basis} \label{sec:GvacGG}
The following is a brief digression deriving~\eqref{molGvacGG}, which
will be beneficial to demonstrate how analytical expressions
for~\eqref{molGvacGG} exist without needing high-dimensional numerical
cubature, and to later extend similar formulas in the particular case
of periodic molecular structures. The derivations below assume
$\omega = \im\xi$, so evaluation at real $\omega$ can be obtained by
substituting $\xi = -\im\omega$ in the results at the end. We define
$G_{0ij} (\im\xi, \vec{x}, \vec{x}') = (\partial_{i} \partial_{j} -
\frac{\xi^{2}}{c^{2}} \delta_{ij}) g_{0} (\im\xi, \vec{x}, \vec{x}')$
where
$g_{0} (\im\xi, \vec{x}, \vec{x}') = \frac{e^{-\xi |\vec{x} -
    \vec{x}'|/c}}{4\pi |\vec{x} - \vec{x}'|}$. This means the inner
product may be written as
$\int \int f_{p} (\vec{x}) (\partial_{i} \partial_{j} - (\xi/c)^{2}
\delta_{ij})g_{0} (\im\xi, \vec{x}, \vec{x}') f_{q}
(\vec{x}')~\mathrm{d}^{3} x'~\mathrm{d}^{3} x$. Performing
integrations by parts given vanishing surface terms to put the
derivatives on $f_{p}$, noting the form of $f_{p}$ as dependent only
on $|\vec{x} - \vec{x}_{p}|$ allows for writing
$\partial_{j} f_{p}(\vec{x}) = -\partial_{r_{pj}} f_{p} (\vec{x})$,
and bringing the derivatives with respect to the Gaussian basis
function centers $\vec{r}_{p}$ outside of the integrals over $\vec{x}$
and $\vec{x}'$ allows for rewriting the inner product as
$(\partial_{r_{pi}}
\partial_{r_{pj}} - (\xi/c)^{2} \delta_{ij}) \iint f_{p} (\vec{x})
g_{0} (\im\xi, \vec{x}, \vec{x}') f_{q} (\vec{x}')~\mathrm{d}^{3}
x'~\mathrm{d}^{3} x$. To simplify this calculation, it is necessary to
write $g_0$ using the following integral representation:
\begin{equation}
  g_{0} (\im\xi, \vec{x}, \vec{x}') = \frac{1}{2\pi^{3/2}}
  \int_{0}^{\infty} \exp(-u^{2} |\vec{x} - \vec{x}'|^{2} -
  (\xi/(2cu))^{2})~\mathrm{d}u,
\end{equation}
in which case the inner product simply turns into a set of Gaussian
integrals. This can be seen in the product, $f_{p}(\vec{x})
g_{0}(i\xi,\vec{x},\vec{x}') f_{q}(\vec{x}')$, whose spatial
dependence comes only in the exponential term, the exponent of which
\begin{equation*}
  \begin{split}
    &-\frac{1}{2\sigma_{p}^{2}} |\vec{x} - \vec{r}_{p}|^{2} - u^{2}
    |\vec{x} - \vec{x}'|^{2} - \frac{1}{2\sigma_{q}^{2}} |\vec{x}' -
    \vec{r}_{q}|^{2} =
    \\ &-\frac{1}{2}\Bigg[\left(\frac{1}{\sigma_{p}^{2}} +
    2u^{2}\right) \vec{x}^{2} - \left(2u^{2} \vec{x} \cdot \vec{x}' +
    2u^{2} \vec{x}' \cdot \vec{x}\right) \\ &+
    \left(\frac{1}{\sigma_{q}^{2}} + 2u^{2}\right)\vec{x}'^{2} \Bigg]
    + \frac{\vec{r}_{p}}{\sigma_{p}^{2}} \cdot \vec{x} +
    \frac{\vec{r}_{q}}{\sigma_{q}^{2}} \cdot \vec{x}' -
    \frac{\vec{r}_{p}^{2}}{2\sigma_{p}^{2}} -
    \frac{\vec{r}_{q}^{2}}{2\sigma_{q}^{2}}
  \end{split}
\end{equation*}
can be written as $-\frac{1}{2} \vec{X}^{\top} \mathbb{A} \vec{X} +
\vec{J}^{\top} \vec{X}$ given the 6-dimensional vectors and tensor,
\begin{align}
  \vec{X} &= \begin{bmatrix} \vec{x} \\ \vec{x}' \end{bmatrix} \\
  \vec{J} &= \begin{bmatrix} \sigma_{p}^{-2} \vec{r}_{p}
    \\ \sigma_{q}^{-2} \vec{r}_{q} \end{bmatrix} \\
  \mathbb{A} &= \begin{bmatrix} (\sigma_{p}^{-2} + 2u^{2})\II &
    -2u^{2}\II \\ -2u^{2} \II & (\sigma_{q}^{-2} +
    2u^{2})\II \end{bmatrix}
\end{align}
Exploiting the well-known identity of multivariate Gaussian
integration,
\begin{equation}
  \begin{split}
    &\int \exp\left(-\frac{1}{2} \vec{X}^{\top} \mathbb{A}
    \vec{X} + \vec{J}^{\top} \vec{X}\right)~\mathrm{d}^{6} X =
    \\ &\frac{(2\pi)^{3}}{\sqrt{\det(\mathbb{A})}}
    \exp\left(\frac{1}{2} \vec{J}^{\top} \mathbb{A}^{-1}
    \vec{J}\right),
  \end{split}
\end{equation}
and accounting for the remaining factors $\frac{1}{2\pi^{3/2}} \times
\frac{1}{(2\pi\sigma_{p} \sigma_{q})^{3}}
\exp\left(-\frac{\vec{r}_{p}^{2}}{2\sigma_{p}^{2}} -
\frac{\vec{r}_{q}^{2}}{2\sigma_{q}^{2}}\right)$ in the multiplication
$f_{p} g_{0} f_{q}$, one obtains
\begin{equation} \label{eq:auxiliaryintegral}
  \begin{split}
    \iint f_{p} &(\vec{x}) g_{0} (\im\xi, \vec{x}, \vec{x}') f_{q}
    (\vec{x}')~\mathrm{d}^{3} x'~\mathrm{d}^{3} x = \\ 
    &\frac{1}{2\pi^{3/2}} \int_{0}^{\infty} (1 + 2(\sigma_{p}^{2} +
    \sigma_{q}^{2})u^{2})^{-3/2} \times
    \\ &\exp\left(-\frac{u^{2} |\vec{r}_{p} -
      \vec{r}_{q}|^{2}}{1 + 2(\sigma_{p}^{2} + \sigma_{q}^{2})u^{2}} -
    \frac{\xi^{2}}{4c^{2} u^{2}}\right)~\mathrm{d}u
  \end{split}
\end{equation}
in terms of an integral over the auxiliary variable $u$. This integral
may be evaluated directly through use of a few variable
substitutions. The first is to transform to dimensionless variables
$v = \sqrt{2(\sigma_{p}^{2} + \sigma_{q}^{2})}u$,
$\rho = \frac{|\vec{r}_{p} - \vec{r}_{q}|}{\sqrt{2(\sigma_{p}^{2} +
    \sigma_{q}^{2})}}$, and
$\theta = \frac{\sqrt{2(\sigma_{p}^{2} + \sigma_{q}^{2})}
  \xi}{c}$. This transforms the integral into
\begin{equation*}
  \frac{1}{(2\pi)^{3/2} \sqrt{\sigma_{p}^{2} + \sigma_{q}^{2}}}
  \int_{0}^{\infty} (1 + v^{2})^{-3/2}
  \exp\left(-\frac{\rho^{2} v^{2}}{1 + v^{2}} -
  \frac{\theta^{2}}{4v^{2}}\right)~\mathrm{d}v
\end{equation*}
such that all dimensional terms are prefactors of the integral, which
itself is dimensionless. The second is to transform to
$w = \frac{v}{\sqrt{1 + v^{2}}}$ so that the semi-infinite integration
range is mapped to the finite interval $[0, 1]$, and
$(1 + v^{2})^{-3/2}~\mathrm{d}v = \mathrm{d}w$, yielding the integral
\begin{equation*}
  \begin{split}
    &\frac{\exp(\theta^{2} / 4)}{(2\pi)^{3/2}
      \sqrt{\sigma_{p}^{2} + \sigma_{q}^{2}}} \int_{0}^{1}
    \exp\left(-\rho^{2} w^{2} -
    \frac{\theta^{2}}{4w^{2}}\right)~\mathrm{d}w =
    \\ &\frac{\exp(\theta^{2} / 4)}{8\pi
      \sqrt{2(\sigma_{p}^{2} + \sigma_{q}^{2})}\rho}
    \left(e^{-\rho\theta} \erfc(\theta/2 - \rho) -
    e^{\rho\theta}\erfc(\theta/2 + \rho)\right)
  \end{split}
\end{equation*}
after direct evaluation. It is this expression that is finally used to
obtain~\eqref{molGvacGG} at real frequency (after substituting
$\xi = -\im\omega$).

\section{Compact Molecules in a Fixed Macroscopic Environment} \label{sec:compact}
For the case of vdW interactions, we start by
considering~\eqref{vdWintegrand} for the case where the macroscopic
bodies (if there are multiple) are fixed relative to each other, so
that $\TT_{\mathrm{mac}} = \TT_{\mathrm{mac}\infty}$. Typically, the
energy differences we choose to measure are set to be relative to a
configuration where the molecules are infinitely separated from the
macroscopic bodies, so the off-diagonal blocks between molecules and
macroscopic bodies in the reference configuration satisfy
$\mathbb{G}_{0\infty} \to 0$. This allows for simplification of the
integrand to
$\ln(\det(\TT_{\mathrm{mol}\infty})\det(\TT_{\mathrm{mol}}^{-1} -
\GG^{(0)} \TT_{\mathrm{mac}} \GG^{(0)}))$. At this point, we may
define
\begin{equation} \label{eq:Gmacdefinition}
  \mathbb{G}^{\mathrm{mac}} = \GG^{(0)} +
  \GG^{(0)} \TT_{\mathrm{mac}} \GG^{(0)}
\end{equation}
as the EM field response in the presence of only the fixed macroscopic
bodies, and
\begin{equation} \label{eq:Tmolmoddefinition}
  \TT_{\mathrm{mol}}'^{-1} = \VV_{\mathrm{mol}}^{-1} -
  \mathbb{G}^{\mathrm{mac}}
\end{equation}
as an effective T-operator encoding the scattering properties among
the molecules in a modified EM environment due to the presence of
fixed macroscopic bodies in the background. This allows for rewriting
the integrand as $\ln(\det(\TT_{\mathrm{mol}\infty}
\TT_{\mathrm{mol}}'^{-1}))$.

For the case of thermal radiation, the definitions
in~\eqref{Gmacdefinition} and~\eqref{Tmolmoddefinition} can only be
used to simplify~\eqref{heattransferonlymolormac} in the case where
the labels $m$ and $n$ are only for molecules; it turns out that a
fuller consideration of macroscopic DOFs is required when at least one
of $m$ or $n$ is a macroscopic body, so~\eqref{Gmacdefinition}
and~\eqref{Tmolmoddefinition} are insufficient in those cases. Thus,
if we focus on the case of heat exchange only among molecules, the
macroscopic bodies again form a fixed background that only act to
modify the field response experienced by the molecules,
so~\eqref{heattransferonlymolormac} is changed to yield
\begin{equation}
  \Phi^{(m)}_{n} = -4~\trace\Big[\asym(\VV_{m}^{-1\dagger}) \PP_{m}
    \TT_{\mathrm{mol}}'^{\dagger} \asym(\PP_{n}
    \mathbb{G}^{\mathrm{mac}}) \TT_{\mathrm{mol}}' \PP_{m}\Big]
\end{equation}
as the thermal energy exchange among molecules.

Our definitions~\eqref{Gmacdefinition} and~\eqref{Tmolmoddefinition}
are useful because $\TT_{\mathrm{mac}}$ is stipulated to be fixed
given that the macroscopic bodies will never change in separation or
orientation relative to each other, so the field response
$\GG^{\mathrm{mac}}$ may be computed using a much broader range of
computational methods, such as finite-difference or multipole methods
in addition to spectral or VIE T-operator methods, in which
$\TT_{\mathrm{mac}}$ by itself may be practically more difficult to
extract. However, this benefit can also be seen as a pitfall in
itself: in almost every situation where $\GG^{\mathrm{mac}}$ can be
computed analytically or numerically, the matrix elements
$\bracket{\vec{f}_{pi}|\GG^{\mathrm{mac}} \vec{f}_{qj}}$ will
generally require slow 6-dimensional numerical cubature, as the
Gaussian widths defining the basis functions are not guaranteed to be
small enough at any given frequency and geometric configuration to be
approximated as point dipoles. The only exception is if
$\GG^{\mathrm{mac}}$ can be analytically written in terms of
$\GG^{(0)}$, as is true, for example, in the case of a PEC plane
thanks to image theory, which can be used as a good approximation for
a thick planar metallic substrate at frequencies below the
ultraviolet; this would allow for using the analytical formulas
of~\eqref{molGvacGG}, avoiding the need for costly numerical
cubature. (An approximate exception, seen in~\figref{pastwork}
(bottom-left), comes from~\cite{VenkataramPRL2017} where we ignored
the effective nuclear DOFs: this yielded much smaller Gaussian basis
function widths, so we approximated $\GG^{\mathrm{mac}} - \GG^{(0)}$
above a gold plate or a gold cone using numerical techniques by
approximating the Gaussian basis functions as Dirac delta
functions. However, this approximation is not valid when nuclear DOFs
are considered.) Thus, our code can currently only treat molecular
bodies interacting either in vacuum or in the presence of a single PEC
plane. This is the implementation we have used for our past
works~\cite{VenkataramPRL2017, VenkataramPRL2018,
  VenkataramSCIADV2019} which primarily consider compact molecules
above a single PEC plane, and we point readers to those works for more
detailed discussions of specific example systems. We also point out
that the Gaussian basis functions $\ket{\vec{f}_{pi}}$ describing the
molecular DOFs do not have compact support, which means those basis
functions will nontrivially overlap with a PEC plane (or equivalently
with their images on the other side of the plane) if present,
potentially leading to unphysical results; this is not problematic for
Gaussian widths smaller than $1~\mathrm{nm}$ as we never consider
smaller separations anyway, while for molecular bodies with large
Gaussian widths especially at smaller separations, even if the results
are not rigorously justifiable, we retain them as a useful
approximation to the short-range EM interaction effects of the
molecular body near a metallic surface.


\begin{figure}[t!]
  \centering
  \includegraphics[width=0.95\columnwidth]{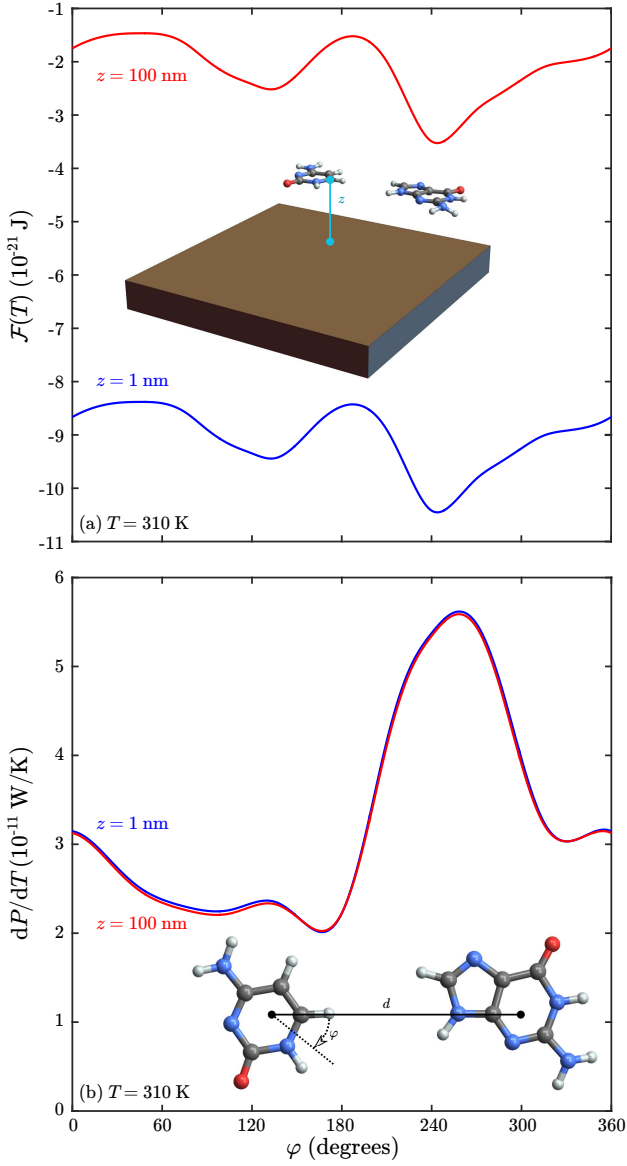}
  \caption{\textbf{Interactions between guanine and cytosine above a
      PEC plane}. (a) vdW interaction free energy $\mathcal{F}(T)$, or
    (b) RHT coefficient $\frac{\mathrm{d}P}{\mathrm{d}T}$, each
    between guanine and cytosine in the presence of a PEC plane. Both
    are at distance $z$ of $1~\mathrm{nm}$ (blue) or $100~\mathrm{nm}$
    (red) above the PEC plane, their centers of mass are displaced
    horizontally by $d = 1~\mathrm{nm}$, and cytosine is rotated
    clockwise about the $z$-axis through its center of mass by angle
    $\varphi$; both calculations are at $T = 310~\mathrm{K}$.}
  \label{fig:guaninecytosinePECvdWRHT}
\end{figure}

As a simple example of interactions involving small compact biological
molecules, in~\figref{guaninecytosinePECvdWRHT}, we consider (a) the
vdW interaction free energy $\mathcal{F}(T)$, or (b) the RHT
coefficient $\frac{\mathrm{d}P}{\mathrm{d}T}$, between the nucleotides
guanine and cytosine as functions of the orientation of cytosine,
given by the clockwise rotation angle $\varphi$ about the $z$-axis
through the center of mass of cytosine, in which the two molecules are
displaced from each other horizontally by a distance $d =
1~\mathrm{nm}$, and both are displaced vertically by the same distance
$z$ above a PEC plane; all calculations are done at the normal human
body temperature $T = 310~\mathrm{K}$. We compare $\mathcal{F}(T)$ and
$\frac{\mathrm{d}P}{\mathrm{d}T}$ for different values of $z$: each
quantity changes by much less than 1\% when $z$ is increased beyond
$100~\mathrm{nm}$, so we only consider $z \in \{1~\mathrm{nm},
100~\mathrm{nm}\}$. The vdW interaction free energy shows clear
differences at each $z$, indicating that there is a significant
contribution from the vertical force by the PEC plane to the overall
interaction for $z \leq 100~\mathrm{nm}$. However, this is completely
independent of $\varphi$, because if the two functions of $\varphi$
are overlaid upon each other to have the same value at $\varphi = 0$,
they consistently remain well within 1\% of each other for all
$\varphi$. This means that the vdW torque $-\frac{\partial
  \mathcal{F}}{\partial \varphi}$ at each $\varphi$ is essentially
independent of $z$; at any $z$, for that value of $d$ and initial
orientation of molecules, there are two stable and two unstable
equilibria for the vdW torque. Meanwhile, the RHT coefficient is
likewise essentially independent of $z$, as is clear from the
figure. The apparent independence of these quantities from the
distance to a PEC plane is due in both cases to consideration of the
interactions between two small, compact, chemically heterogeneous
molecules of complicated shapes, yielding weaker polarization
responses, as opposed to the interactions between a low-dimensional
compact or extended low-dimensional carbon allotropes of simple
high-symmetry shapes, which would yield stronger polarization
responses. This has previously been observed in comparisons between
the interactions with a metal plate of low-dimensional carbon
allotropes versus complicated proteins~\cite{VenkataramPRL2017}.

\begin{figure}[t!]
  \centering
  \includegraphics[width=0.95\columnwidth]{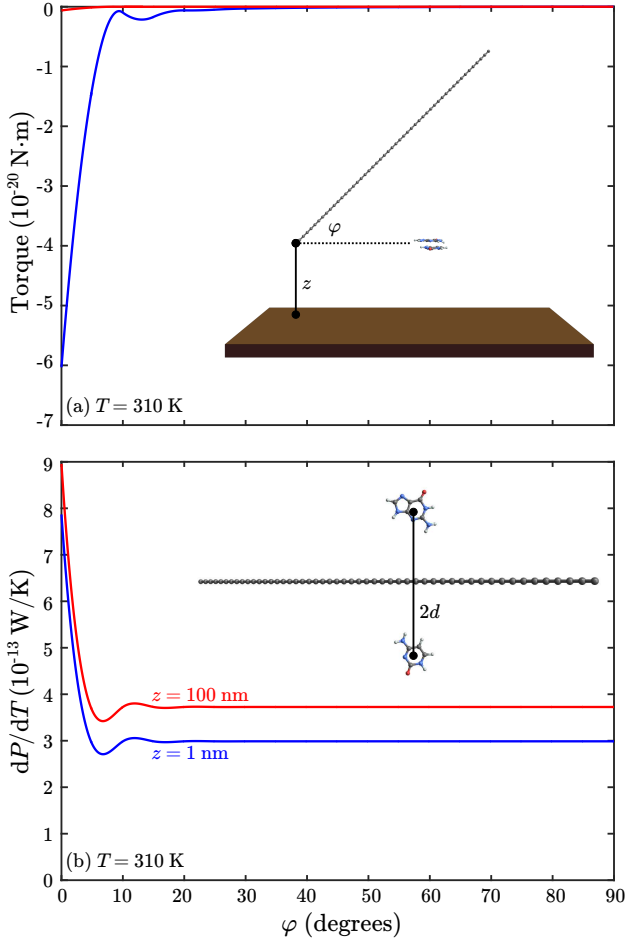}
  \caption{\textbf{Interactions among guanine, cytosine, and a 250
      atom-long carbyne wire above a PEC plane}. (a) vdW interaction
    torque, or (b) RHT coefficient
    $\frac{\mathrm{d}P}{\mathrm{d}T}$. In both calculations, the
    bottom atom of the wire and the centers of mass of guanine and
    cytosine are at a distance $z$, namely $1~\mathrm{nm}$ (blue) or
    $100~\mathrm{nm}$ (red), above a PEC plane, and the wire may be
    oriented with respect to the horizontal axis at an angle
    $\varphi$. Furthermore, at $\varphi = 0$ (corresponding to the
    wire lying parallel to the PEC plane), the centers of mass of the
    wire, guanine, and cytosine lie in a line perpendicular to the
    wire axis and parallel to the PEC plane, with guanine and cytosine
    each lying at $d = 1~\mathrm{nm}$ on opposite sides of the
    projection of the wire onto the horizontal plane. Both
    calculations are at $T = 310~\mathrm{K}$. Top inset: schematic of
    the carbyne wire at an angle $\varphi$ above the PEC plane. Bottom
    inset: plan (top) view of the wire, guanine, and cytosine. In both
    schematics, the wire is not to scale, but everything else is to
    scale.}
  \label{fig:guaninecytosinecarbynePECvdWRHT}
\end{figure}

As a more complex illustrative example leading to nontrivial
interaction behaviors, in~\figref{guaninecytosinecarbynePECvdWRHT}, we
consider (a) the vdW interaction torque, or (b) the RHT coefficient
$\frac{\mathrm{d}P}{\mathrm{d}T}$, of a complicated four-body system
involving guanine \& cytosine along with a 250 atom-long carbyne wire,
all of which lie above a PEC plane. The two small molecules' centers
of mass, as well as the bottom of the carbyne wire, are fixed at a
distance $z$ above the PEC plane, while the angle $\varphi$ that the
wire makes with respect to the horizontal plane is varied. At $\varphi
= 0$, the wire comes exactly in between the two small molecules, such
that the centers of mass of these three molecules lie along a line
perpendicular to the wire, with each small molecule's center of mass a
distance $2d = 2~\mathrm{nm}$ away from the other. Effectively, the
wire can be thought of as a ``switch'' to be lifted, as we consider
variations in the torque and heat transfer coefficient with respect to
the orientation $\varphi$ for two different values of $z$, namely $z =
1~\mathrm{nm}$ or $z = 100~\mathrm{nm}$. All of these quantities are
computed at the normal human body temperature $T = 310~\mathrm{K}$.

The behavior of the torque is dominated by the interactions of the
wire with the PEC plane, and guanine and cytosine, being small
molecules with no obvious symmetries \& weak polarizabilities, only
have a marginal impact on the interaction torque. In particular, the
configuration at $z = 100~\mathrm{nm}$ is far enough from the PEC
plane that the torque is always very close to zero, as an isolated
carbyne wire in free space should not exhibit any torque at all; for
$\varphi \geq 15~\mathrm{degrees}$ (i.e. even when the wire is
somewhat close to parallel to the PEC plane), the wire is far enough
from the small molecules that the torque is effectively negligible,
while it only has a magnitude of $6.2\times 10^{-22}~\mathrm{N\cdot
  m}$ (with the negative sign indicating attraction to the PEC plane
and other molecules) even at the parallel orientation $\varphi =
0$. By contrast, at $z = 1~\mathrm{nm}$, the proximate PEC plane
produces a much stronger attraction even at relatively large $\varphi$
(i.e. when the wire is closer to perpendicular to the PEC plane), and
the attractive torque of $6 \times 10^{-20}~\mathrm{N\cdot m}$ at
$\varphi = 0$ is nearly 2 orders of magnitude larger than the
corresponding torque at $z = 100~\mathrm{nm}$.

Qualitatively subtler effects emerge when considering the heat
transfer coefficient specifically between guanine and cytosine,
because in this situation, the wire and the PEC plane both act only to
modify the environmental EM scattering properties, with the wire
orientation angle $\varphi$ further varying this; thus, aggregate
many-body effects must be considered. At $z = 100~\mathrm{nm}$, the
wire being parallel to the PEC plane and lying between the two small
molecules facilitates heat transfer between the two small molecules
compared to when it is perpendicular to the PEC plane (such that the
small molecules are exchanging energy effectively in vacuum), by
virtue of modifying the EM scattering properties: the heat transfer
coefficient decreases from $8.96\times 10^{-13}~\mathrm{W/K}$ at
$\varphi = 0$ to $3.73\times 10^{-13}~\mathrm{W/K}$ at $\varphi =
90~\mathrm{degrees}$. Qualitatively similar behavior is observed at $z
= 1~\mathrm{nm}$, but screening from the proximate PEC plane
(i.e. from the image dipoles of each molecule) consistently decreases
the heat transfer coefficients, and this decrease is not uniform with
respect to $\varphi$, as the ratio of the heat transfer coefficient at
$z = 1~\mathrm{nm}$ to its counterpart at $z = 100~\mathrm{nm}$
decreases nonmonotonically from $0.88$ at $\varphi = 0$ to $0.8$ at
$\varphi = 90~\mathrm{degrees}$.

While our past works~\cite{VenkataramPRL2017, VenkataramPRL2018,
  VenkataramSCIADV2019} have focused on power laws or ratios of vdW
interaction free energies or RHT powers with respect to distances, and
have focused on high-symmetry low-dimensional carbon allotropes
present individually or in pairs in vacuum or near a PEC plane, here
we show the greater generality of our framework in treating compact
molecules, modeling FED phenomena in more complex many-body systems of
carbon allotropes and biological molecules in the presence of a PEC
plane, and considering rotational dependence in addition to distance
dependence. We expect that our framework, as given, may have fruitful
applications to predictions of highly nontrivial vdW interaction
energy and RHT power behaviors in even more complex systems of compact
molecules near metallic substrates as experiments become more
sensitive.

\section{Extended Molecular Structures in an Arbitrary Environment} \label{sec:periodic}

In this section, we extend expressions from the previous sections for
the vdW interaction energy and heat exchange among a collection of
molecular and macroscopic bodies to consider spatially extended
geometries with commensurate spatial periodicity. The imposition of
Bloch periodicity leads to nontrivial expressions of the
fluctuation--dissipation theorem as well as novel formulas
extending~\eqref{molGvacGG} to spatially extended structures. Given
this, we first review the definitions and relations among EM fields,
polarizations, and response functions in periodic geometries, and
derive relevant FED formulas taking care with the nontrivial changes
imposed by Bloch periodicity in~\Secref{periodicformulas}. Then, we
exploit Ewald summations and integral techniques to derive
fast-converging semianalytical expressions for the matrix elements
that describe scattering among periodic molecular basis functions in
vacuum in~\Secref{Ewald}. Finally, we consider examples of relevant
atomistic systems exhibiting periodic boundary conditions to which
this formalism may be applied, particularly vdW interactions and
RHT between two parallel graphene sheets in
vacuum, in~\Secref{graphene} as a demonstration of the versatility of
our method.

\subsection{Scattering, vdW interactions, and thermal radiation
  among polarizable bodies with periodic
  boundaries} \label{sec:periodicformulas}

Consider a collection of polarizable bodies labeled $n \in \{1,
\ldots, N\}$ that all obey periodic boundary conditions given by
lattice vectors $\vec{R}$; these polarizable bodies will first be
treated in a fully general manner (without regard to whether they
should be treated atomistically or continuously), and are assumed to
be disjoint, such that $\VV = \sum_{n = 1}^{N} \VV_{n}$. The
periodicity of this system allows for writing this more explicitly in
terms of projections at each unit cell. Denoting $\PP_{\vec{R}}$ as
the projection operator onto the subspace spanned by the DOFs in the
unit cell at $\vec{R}$, the resolution of the identity can be written
as $\II = \sum_{\vec{R}} \PP_{\vec{R}}$. Further defining
$\ket{\vec{P}_{\vec{R}}} = \PP_{\vec{R}} \ket{\vec{P}}$ (and likewise
$\ket{\vec{E}_{\vec{R}}} = \PP_{\vec{R}} \ket{\vec{E}}$) allows for
writing $\ket{\vec{P}_{\vec{R}}} = \sum_{\vec{R}'}
\VV_{\vec{R},\vec{R}'} \ket{\vec{E}_{\vec{R}'}}$, where
$\VV_{\vec{R},\vec{R}'} = \PP_{\vec{R}} \VV \PP_{\vec{R}'}$.


At this point, it is useful to expand the real-space representations
of these quantities in terms of Bloch-periodic functions within each
unit cell. This means expressing
\begin{equation}
  \ket{\vec{P}_{\vec{k}}} = \sum_{\vec{R}} e^{-\im\vec{k} \cdot
    \vec{R}} \ket{\vec{P}_{\vec{R}}}
\end{equation}
and its inverse,
$
  \ket{\vec{P}_{\vec{R}}} = \frac{V_{\mathrm{uc}}}{(2\pi)^{d}}
  \int_{\mathrm{BZ}} \ket{\vec{P}_{\vec{k}}} e^{\im\vec{k} \cdot
    \vec{R}}~\mathrm{d}^{d} k,
$
(with similar expressions for $\ket{\vec{E}}$) in terms of the Bloch
wavevector $\vec{k}$, which is assumed to lie within the first
Brillouin zone (BZ) which has volume $(2\pi)^{d} / V_{\mathrm{uc}}$
given in terms of the unit cell volume $V_{\mathrm{uc}}$. These come
from the completeness relations
$
  \sum_{\vec{R}} e^{-\im\vec{k} \cdot \vec{R}} =
  \frac{(2\pi)^{d}}{V_{\mathrm{uc}}} \delta^{d} (\vec{k})
$
in real space and similarly
$
  \frac{V_{\mathrm{uc}}}{(2\pi)^{d}} \int_{\mathrm{BZ}} e^{\im\vec{k}
    \cdot \vec{R}}~\mathrm{d}^{d} k = \delta_{\vec{R},0}
$
in reciprocal space.

The susceptibility $\VV_{\vec{R},\vec{R}'}$, by virtue of representing
a periodic system, has translational symmetry across unit cells, so
$\VV_{\vec{R}, \vec{R}'} = \VV_{\vec{R} - \vec{R}',0}$; alternatively,
$\VV_{\vec{R} + \vec{R}'', \vec{R}' + \vec{R}''} = \VV_{\vec{R},
  \vec{R}'}$. Since the susceptibility $\VV_{\vec{R}, 0}$ is the
polarization response within the unit cell centered at $\vec{R}$ to an
electric field applied to atoms in unit cell 0, namely
$\delta_{\vec{R}, 0} \II$, its reciprocal space representation
is given by:
\begin{equation}
  \VV_{\vec{k}} = \sum_{\vec{R}} e^{-\im\vec{k} \cdot \vec{R}}
  \VV_{\vec{R}, 0}
\end{equation}
where the choice of unit cell 0 is arbitrary due to the discrete
translational symmetry underlying this system; namely, changing the
summand $\VV_{\vec{R}, 0}$ to $\VV_{\vec{R}, \vec{R}'}$ changes
$\VV_{\vec{k}}$ to $e^{-\im\vec{k} \cdot \vec{R}'} \VV_{\vec{k}}$,
reflecting the Bloch periodicity of the response. Additionally,
reciprocity $\VV = \VV^{\top}$ in position space implies that
$\VV_{\vec{R}, 0} = (\VV_{0, \vec{R}})^{\top}$, from which it follows
that $(\VV_{\vec{k}})^{\top} = \VV_{-\vec{k}}$ in Bloch space. Hence,
the relationship between the polarization and electric field in
reciprocal space is
\begin{equation}
  \ket{\vec{P}_{\vec{k}}} = \VV_{\vec{k}} \ket{\vec{E}_{\vec{k}}},
\end{equation}
and the transformation to reciprocal space partially diagonalizes the
problem, reducing it to one that can be solved within the unit cell.

All of the above relations also hold when considering the vacuum
electromagnetic field Green's function $\GG^{(0)}$ relating
$\ket{\vec{E}} = \GG^{(0)} \ket{\vec{P}}$ and solving
Maxwell's equations in vacuum
\begin{equation}
  \Big[(c/\omega)^{2} \nabla \times (\nabla \times) - \II\Big]
  \GG^{(0)} = \II,
\end{equation}
under the \emph{same} periodicity as the 
susceptibility $\VV$. Thus, the Green's function in reciprocal space
can be written as
\begin{equation}
  \GG^{(0)}_{\vec{k}} = \sum_{\vec{R}} e^{-\im\vec{k}
    \cdot \vec{R}} \GG^{(0)}_{\vec{R}, 0}
\end{equation}
for $\vec{k}$ in the BZ, and $\ket{\vec{E}_{\vec{k}}} =
\GG^{(0)}_{\vec{k}} \ket{\vec{P}_{\vec{k}}}$.

Given source polarizations $\ket{\vec{P}^{(0)}_{\vec{R}}}$ and fields
$\ket{\vec{E}^{(0)}_{\vec{R}}}$, Maxwell's equations in integral form
can be written as
\begin{align} \label{eq:Maxwellintegralperiodic}
  \ket{\vec{P}_{\vec{R}}} &= \ket{\vec{P}^{(0)}_{\vec{R}}} +
  \sum_{\vec{R}'} \VV_{\vec{R},\vec{R}'} \ket{\vec{E}_{\vec{R}'}}
  \\
  \ket{\vec{E}_{\vec{R}}} &= \ket{\vec{E}^{(0)}_{\vec{R}}} +
  \sum_{\vec{R}'} \GG^{(0)}_{\vec{R},\vec{R}'}
  \ket{\vec{P}_{\vec{R}'}}
\end{align}
for this system. These equations
become easier to manipulate in reciprocal space. In particular,
~\eqref{Maxwellintegralperiodic} becomes
\begin{align}
  \ket{\vec{P}_{\vec{k}}} &= \ket{\vec{P}^{(0)}_{\vec{k}}} +
  \VV_{\vec{k}} \ket{\vec{E}_{\vec{k}}} \\
  \ket{\vec{E}_{\vec{k}}} &= \ket{\vec{E}^{(0)}_{\vec{k}}} +
  \GG^{(0)}_{\vec{k}} \ket{\vec{P}_{\vec{k}}}
\end{align}
for $\vec{k}$ in the BZ, so these can be formally solved to yield
\begin{align}
  \ket{\vec{P}_{\vec{k}}} &= \TT_{\vec{k}}
  \left(\VV^{-1}_{\vec{k}} \ket{\vec{P}^{(0)}_{\vec{k}}} +
  \ket{\vec{E}^{(0)}_{\vec{k}}}\right) \\ \ket{\vec{E}_{\vec{k}}} &=
  (\II + \GG^{(0)}_{\vec{k}}
  \TT_{\vec{k}})\ket{\vec{E}^{(0)}_{\vec{k}}} +
  \GG^{(0)}_{\vec{k}} \TT_{\vec{k}}
  \VV^{-1}_{\vec{k}} \ket{\vec{P}^{(0)}_{\vec{k}}}
\end{align}
just as in~\eqref{maxwellintegralsolve}, where $\TT_{\vec{k}} =
(\VV^{-1}_{\vec{k}} - \GG^{(0)}_{\vec{k}})^{-1}$.

If the free polarization sources and incident fields arise from
quantum and thermal fluctuations, they satisfy the
fluctuation--dissipation theorem
\begin{equation}
  \begin{split}
    \bracket{\ket{\vec{P}^{(0)}_{\vec{R}} (\omega)}
      \bra{\vec{P}^{(0)}_{\vec{R}'} (\omega')}} &=
    \frac{2\Theta(\omega, T)}{\omega} \Im(\VV_{\vec{R}, \vec{R}'})
    \times 2\pi\delta(\omega - \omega')
    \\ \bracket{\ket{\vec{E}^{(0)}_{\vec{R}} (\omega)}
      \bra{\vec{E}^{(0)}_{\vec{R}'} (\omega')}} &=
    \frac{2\Theta(\omega, T)}{\omega} \Im(\GG^{(0)}_{\vec{R},
      \vec{R}'}) \times 2\pi\delta(\omega - \omega')
  \end{split}
\end{equation}
after exploiting reciprocity to equate $\asym(\VV) = \Im(\VV)$ and
$\asym(\GG^{(0)}) = \Im(\GG^{(0)})$ in position space. At this point,
it becomes necessary to transform the fluctuation--dissipation
theorems into Bloch space. As the fluctuating fields and polarizations
are correlated only with themselves and not with each other, then
$\bracket{\ket{\vec{E}^{(0)}_{\vec{R}} (\omega)}
  \bra{\vec{P}^{(0)}_{\vec{R}'} (\omega')}} = 0$ as before. We start
with
\begin{multline} \label{eq:hybridFDT}
    \bracket{\ket{\vec{P}^{(0)}_{\vec{k}} (\omega)}
      \bra{\vec{P}^{(0)}_{\vec{k}'} (\omega')}} = \\ \sum_{\vec{R},
      \vec{R}'} e^{-\im (\vec{k} \cdot \vec{R} - \vec{k}' \cdot
      \vec{R}')} \bracket{\ket{\vec{P}^{(0)}_{\vec{R}} (\omega)}
      \bra{\vec{P}^{(0)}_{\vec{R}'} (\omega')}} = \\
    \frac{2\Theta(\omega, T)}{\omega} \sum_{\vec{R}, \vec{R}'} e^{-\im
      (\vec{k} \cdot \vec{R} - \vec{k}' \cdot \vec{R}')}
    \Im(\VV_{\vec{R}, \vec{R}'}) \times 2\pi\delta(\omega - \omega')
\end{multline}
from using the real space fluctuation--dissipation theorem. If both
sides are integrated over $\vec{k}'$ in the BZ, then this yields
\begin{multline}
  \frac{V_{\mathrm{uc}}}{(2\pi)^{d}} \int_{\mathrm{BZ}}
  \bracket{\ket{\vec{P}^{(0)}_{\vec{k}} (\omega)}
    \bra{\vec{P}^{(0)}_{\vec{k}'} (\omega')}}~\mathrm{d}^{d} k' =
  \\ \frac{2\Theta(\omega, T)}{\omega} \sum_{\vec{R}} e^{-\im \vec{k}
    \cdot \vec{R}} \Im(\VV_{\vec{R}, 0}) \times 2\pi\delta(\omega -
  \omega')
\end{multline}
using the reciprocal space relation
$\frac{V_{\mathrm{uc}}}{(2\pi)^{d}} \int_{\mathrm{BZ}} e^{\im \vec{k}'
  \cdot \vec{R}'}~\mathrm{d}^{d} k' =
\delta_{\vec{R}', 0}$. Additionally, as $\Im(\VV_{\vec{R}, 0}) =
(\VV_{\vec{R}, 0} - \VV^{\star}_{\vec{R}, 0})/(2\im)$, then
$\sum_{\vec{R}} e^{-\im \vec{k} \cdot \vec{R}}
\Im(\VV_{\vec{R}, 0}) = (2\im)^{-1} (\VV_{\vec{k}} -
\sum_{\vec{R}} e^{-\im \vec{k} \cdot \vec{R}}
\VV^{\star}_{\vec{R}, 0})$. The second term can be evaluated as
$\sum_{\vec{R}} e^{-\im \vec{k} \cdot \vec{R}}
\VV^{\star}_{\vec{R}, 0} = (\sum_{\vec{R}} e^{\im \vec{k} \cdot
  \vec{R}} \VV_{\vec{R}, 0})^{\star} = \VV^{\star}_{-\vec{k}} =
\VV^{\dagger}_{\vec{k}}$, so this finally yields the integrated
reciprocal space fluctuation--dissipation theorem
$  
\frac{V_{\mathrm{uc}}}{(2\pi)^{d}} \int_{\mathrm{BZ}}
  \bracket{\ket{\vec{P}^{(0)}_{\vec{k}} (\omega)}
    \bra{\vec{P}^{(0)}_{\vec{k}'} (\omega')}}~\mathrm{d}^{d} k' =
  \frac{2\Theta(\omega, T)}{\omega} \asym(\VV_{\vec{k}}) \times 2\pi\delta(\omega - \omega')
$
which in turn yields the Bloch space fluctuation--dissipation theorem:
\begin{multline}
  \bracket{\ket{\vec{P}^{(0)}_{\vec{k}} (\omega)} \bra{\vec{P}^{(0)}_{\vec{k}'} (\omega')}}
    = \frac{2\Theta(\omega, T)}{\omega} \asym(\VV_{\vec{k}} (\omega)) \times \\ 2\pi\delta(\omega - \omega')
    \frac{(2\pi)^{d}}{V_{\mathrm{uc}}} \delta^{d} (\vec{k} - \vec{k}').
\end{multline}
As the same reciprocity properties of $\VV_{\vec{k}}$ hold for
$\GG^{(0)}_{\vec{k}}$, then it also follows that (at thermal
equilibrium),
\begin{multline}
  \bracket{\ket{\vec{E}^{(0)}_{\vec{k}} (\omega)} \bra{\vec{E}^{(0)}_{\vec{k}'}} (\omega')}
  = \frac{2\Theta(\omega, T)}{\omega}
  \asym(\GG^{(0)}_{\vec{k}} (\omega)) \times \\ 2\pi\delta(\omega -
  \omega') \frac{(2\pi)^{d}}{V_{\mathrm{uc}}} \delta^{d} (\vec{k} -
  \vec{k}').
\end{multline}

Having thus derived the fluctuation--dissipation theorems for systems
with Bloch periodicity, we may now derive the vdW free energy at
equilibrium temperature $T$. This once again requires evaluation of
the quantity $\bracket{\bracket{\vec{P}(\lambda), \vec{E}(\lambda))}}
=
\bracket{\trace(\ket{\vec{E}(\lambda)}\bra{\vec{P}(\lambda)})}$. Taking
$\lambda = 1$ for now (restoring explicit factors of $\lambda$ later),
$\bracket{\trace[\ket{\vec{E} (\omega)}\bra{\vec{P} (\omega')}]} =
\sum_{\vec{R}} \bracket{\trace[\ket{\vec{E}_{\vec{R}}
      (\omega)}\bra{\vec{P}_{\vec{R}} (\omega')}]}$. Using the fact
that,
\begin{multline*}
  \bracket{\trace[\ket{\vec{E}_{\vec{R}}
        (\omega)}\bra{\vec{P}_{\vec{R}} (\omega')}]} =
  (V_{\mathrm{uc}}/(2\pi)^{d})^{2}\times \\ \sum_{\vec{R}}
  \int_{\mathrm{BZ}} \int_{\mathrm{BZ}} e^{\im(\vec{k} -
    \vec{k}')\cdot \vec{R}} \bracket{\trace[\ket{\vec{E}_{\vec{k}}
        (\omega)}\bra{\vec{P}_{\vec{k'}} (\omega')}]}~\mathrm{d}^{d}
  k~\mathrm{d}^{d} k'
\end{multline*}
and that,
\begin{multline}
  \bracket{\trace[\ket{\vec{E}_{\vec{k}}
        (\omega)}\bra{\vec{P}_{\vec{k'}} (\omega')}]} =
  \frac{2\Theta(\omega, T)}{\omega}
  \trace[\asym(\GG^{(0)}_{\vec{k}} \TT_{\vec{k}})] \times
  \\ 2\pi\delta(\omega - \omega') \frac{(2\pi)^{d}}{V_{\mathrm{uc}}}
  \delta^{d} (\vec{k} - \vec{k}')
\end{multline}
in analogy to the case of compact molecules, after using the
fluctuation--dissipation theorems for periodic structures and the
properties of Dirac delta functions, the rest of the derivation
follows essentially identically to the case of compact polarizable
bodies. The integration of the Dirac delta function leaves a sum over
$\vec{R}$ of a quantity independent of $\vec{R}$; this physically
reflects the invariance of this periodic problem with respect to
discrete translations, and the fact that periodic structures are
infinite, albeit with the interaction free energy per unit cell
remaining finite. Ultimately, the vdW interaction free energy among a
collection of extended polarizable bodies, per unit cell, is given by:
\begin{equation}
  \mathcal{F}_{\mathrm{uc}} = k_{\mathrm{B}} T\sum_{l = 0}^{\infty}~'
  \int_{\mathrm{BZ}} \ln(\det(\TT_{\infty,\vec{k}}
  \TT^{-1}_{\vec{k}}))~\frac{V_{\mathrm{uc}} \mathrm{d}^{d}
    k}{(2\pi)^{d}}
\end{equation}
where the prime again implies a half weight on the $l = 0$ term in the
sum.


One can also derive a compact formula for the radiation spectrum per
unit cell between extended polarizable bodies by following the same
steps as in the case of compact molecules. In particular, the
radiation spectrum between polarizable bodies $m$ and $n$ (which may
be the same) at a given $\omega$ can be written as
\begin{widetext}
\begin{equation}
  \Phi^{(m)}_{n} = -4\int_{\mathrm{BZ}}
  \trace\left[\asym(\VV_{m,\vec{k}}^{-1 \dagger})\PP_{m}
    \TT^{\dagger}_{\vec{k}} \asym(\PP_{n} \GG^{(0)}_{\vec{k}})
    \TT_{\vec{k}} \PP_{m}\right]~\frac{V_{\mathrm{uc}} \mathrm{d}^{d}
    k}{(2\pi)^{d}}
\end{equation}
\end{widetext}
where dependence on $(\omega, \vec{k})$ is made implicit. The formulas
for the far-field emission $W^{(m)}$ and the heat transfer $W_{m \to
  n}$ remain the same in terms of $\Phi^{(m)}_{n}$, though they are
now thermal emission or RHT spectra per unit cell.

\subsection{Fast molecular scattering matrix elements evaluations via
   Ewald summation} \label{sec:Ewald}

The above formulas do not make explicit reference to molecular or
macroscopic bodies, but just as for general polarizable bodies, the
corresponding DOFs may be separated to yield formulas that yield
greater physical insight and ease of implementation. In particular,
the molecular basis functions are the same as in the compact case, and
just as for general polarizable bodies, if the configuration of Bloch
periodic macroscopic bodies is fixed, they form a scattering
background with a modified Green's function in which vdW interactions
and radiative energy exchange may be computed among periodic molecular
structures. In practice, just as in the compact case, for most
macroscopic geometries, the matrix elements
$\bracket{\vec{f}_{pi}|\GG^{\mathrm{mac}} \vec{f}_{qj}}$ would need to
be computed using costly 6-dimensional numerical cubature; Bloch
periodic boundary conditions adds another cost in the form of summing
over lattice vectors too. For this reason, our code only implements
computations where $\GG^{\mathrm{mac}}$ can be expressed analytically
in terms of $\GG^{(0)}$, namely when either no macroscopic body is
present (i.e. vacuum) or a single PEC plane is present (which can
again be computed via image theory). This may not be such a severe
practical limitation though, as materials like graphene, which have
become of great recent scientific interest, can be treated
atomistically in our model, so interactions among graphene sheets and
molecular crystals in vacuum may be considered without significant
issues. Additionally, the same caveats as for compact molecular
structures apply with respect to the overlaps of the Gaussian basis
functions with a PEC plane. With this in mind, we now turn to deriving
the expressions for the vacuum Green's function matrix elements in the
molecular basis in periodic geometries. The expression of the vacuum
Green's function in terms of~\eqref{auxiliaryintegral} and the
facility in \emph{analytically} performing the resulting spatial
integrals over Gaussian basis functions ensures that the formulas we
obtain are analytical and fast converging over the real and reciprocal
lattice summations; the expressions bear many similarities with Ewald
summation, while the nonzero Gaussian widths ensure that certain
divergences are mitigated, just as for isolated (non-periodic) basis
functions. We perform the following derivations at $\omega = \im\xi$,
and notationally suppress the functional dependence on $\omega$ for
brevity; formulas valid for real $\omega$ can be obtained by
substituting $\xi = -\im\omega$ at the end results.


Our use of Gaussian basis functions of relatively large widths
(especially so when one considers phonons~\cite{VenkataramPRL2018,
  VenkataramSCIADV2019}), ensures that in periodic geometries, the
field responses $\GG^{(0)}_{\vec{k}}$ can no longer be treated from
the perspective of simple point dipoles. Instead, one must directly
compute the matrix elements
\begin{equation*}
  G^{(0)}_{\vec{k}pi,qj} = \sum_{\vec{R}} e^{-\im\vec{k} \cdot
    \vec{R}} \bracket{\vec{f}_{p+\vec{R}, i}|\GG^{(0)} \vec{f}_{qj}}
\end{equation*}
using the definitions of the basis functions $\ket{\vec{f}_{pi}}
\equiv \ket{f_{p} \vec{e}_{i}}$, where the widths of the Gaussian
basis functions $f_{p}$ depend on $(\im\xi, \vec{k})$ via the
susceptibility matrix $\alpha_{\vec{k}}$; notationally,
$\ket{\vec{f}_{p+\vec{R}, i}}$ refers to the periodic image of
$\ket{\vec{f}_{pi}}$ at lattice vector $\vec{R}$, and is represented
in position space as $f_{p}(\vec{x} - \vec{R})
\vec{e}_{i}$. Performing this summation over the real lattice yields
slow conditional convergence, so the goal is to transform this
summation into equivalent fast and absolutely convergent sums,
accounting for the nontrivial Gaussian screening widths. In
particular, this involves rewriting
\begin{multline}
  G^{(0)}_{\vec{k}pi, qj} =
  (\partial_{r_{pi}} \partial_{r_{pj}} - (\xi/c)^{2} \delta_{ij})
  \times \\ \sum_{\vec{R}} e^{-\im\vec{k} \cdot \vec{R}} \iint
  f_{p} (\vec{x} - \vec{R}) g_{0} (\im\xi, \vec{x}, \vec{x}') f_{q}
  (\vec{x}')~\mathrm{d}^{3} x'~\mathrm{d}^{3} x
\end{multline}
and then splitting the integral in eq.~\eqref{auxiliaryintegral} over
$u$, from $[0, \infty)$ to the ranges $[0, \kappa)$ and
$[\kappa, \infty)$, where $\kappa$ is a user-specified Ewald splitting
parameter that controls the speed of
convergence~\cite{GallinetJOSAA2010, CapolinoJCP2007}. Explicitly, this
involves writing
\begin{equation}
  G^{(0)}_{\vec{k}pi,qj} = G^{(0)\mathrm{LR}}_{\vec{k}pi,qj} + G^{(0)\mathrm{SR}}_{\vec{k}pi,qj}
\end{equation}
such that $G^{(0)\mathrm{LR}}_{\vec{k}pi,qj}$ corresponds to
integration over $u \in [0, \kappa)$, while
  $G^{(0)\mathrm{SR}}_{\vec{k}pi,qj}$ corresponds to integration over
  $u \in [\kappa, \infty)$. Our derivations of
    $G^{(0)\mathrm{LR}}_{\vec{k}pi,qj}$ and
    $G^{(0)\mathrm{SR}}_{\vec{k}pi,qj}$ for systems with periodicity
    in 1 or 2 dimensions follow~\cite{GallinetJOSAA2010,
      CapolinoJCP2007}, but with appropriate changes accounting for
    the molecular basis functions having a finite Gaussian spread
    rather than corresponding to point dipoles.

The term $G^{(0)\mathrm{SR}}_{\vec{k}pi,qj}$ is evaluated over the
real lattice, giving expressions independent of periodic
dimensionality. In particular, making the same variable substitutions
$v = \sqrt{2(\sigma_{p}^{2} + \sigma_{q}^{2})}u$ and $w = v/\sqrt{1 +
  v^{2}}$, along with $\mu = \sqrt{2(\sigma_{p}^{2} +
  \sigma_{q}^{2})}\kappa$, $\nu = \mu/\sqrt{1 + \mu^{2}}$, $\rho =
|\vec{r}_{p} + \vec{R} - \vec{r}_{q}|/\sqrt{2(\sigma_{p}^{2} +
  \sigma_{q}^{2})}$, and $\theta = \sqrt{2(\sigma_{p}^{2} +
  \sigma_{q}^{2})}\xi/c$, then carrying out the integration with
respect to $w$ over the range $[\nu, 1)$ yields
\begin{widetext}
\begin{multline*}
    G^{(0)\mathrm{SR}}_{\vec{k}pi,qj} =
    (\partial_{r_{pi}} \partial_{r_{pj}} - (\xi/c)^{2} \delta_{ij})
    \times \sum_{\vec{R}} \frac{e^{(\sigma_{p}^{2} +
        \sigma_{q}^{2})\xi^{2} / (2c^{2}) - \im\vec{k} \cdot
        \vec{R}}}{8\pi |\vec{x}_{p} + \vec{R} - \vec{x}_{q}|} \times
    \\ \Bigg\{e^{-\rho\theta} \left[\erfc\left(\nu\rho -
    \frac{\theta}{2\nu}\right) - \erfc\left(\rho -
    \frac{\theta}{2}\right)\right] + e^{\rho\theta}
    \left[\erfc\left(\nu\rho + \frac{\theta}{2\nu}\right) -
    \erfc\left(\rho + \frac{\theta}{2}\right)\right]\Bigg\}
\end{multline*}
\end{widetext}
for any periodic lattice.

The term $G^{(0)\mathrm{LR}}_{\vec{k}pi,qj}$ is evaluated over the
reciprocal lattice, leading to different expressions for different
periodic dimensionalities. For a 1D-periodic system, the lattice
vectors lie along a single direction with $\vec{R} = n\vec{a}$ (where
$a = |\vec{a}|$), and the reciprocal lattice vectors are likewise
$\vec{g} = n\vec{b}$, where $\vec{b} = 2\pi \vec{a}/a^{2}$. Defining
$\vec{r}_{p} - \vec{r}_{q} = \Delta r_{\parallel} \vec{a}/a + \Delta
\vec{r}_{\perp}$ where $\Delta r_{\parallel}$ is the component of the
displacement between the two atoms along the periodic axis $\vec{a}$
and $\Delta \vec{r}_{\perp}$ is the orthogonal projection, then in the
integrand, $|\vec{r}_{p} + \vec{R} - \vec{r}_{q}|^{2} = (\Delta
r_{\parallel} + na)^{2} + \Delta \vec{r}_{\perp}^{2}$. The real
lattice sum is expressed as:
\begin{multline*}
    \sum_{n = -\infty}^{\infty} \frac{e^{\theta^{2} / 4 - \im\vec{k}
        \cdot \vec{R}}}{2\pi^{3/2} \sqrt{2(\sigma_{p}^{2} +
        \sigma_{q}^{2})}} \times \\\int_{0}^{\nu}
    \exp\left(-\frac{w^{2} (\Delta r_{\parallel} + na)^{2} +
      \Delta \vec{r}_{\perp}^{2}}{2(\sigma_{p}^{2} + \sigma_{q}^{2})}
    - \frac{\theta^{2}}{4w^{2}}\right)~\mathrm{d}w.
\end{multline*}
Defining the function,
\begin{multline*}
    f(l) = \frac{e^{\theta^{2} / 4 - \im kl}}{2\pi^{3/2}
      \sqrt{2(\sigma_{p}^{2} + \sigma_{q}^{2})}} \times
    \\\exp\left(-\frac{\theta^{2}}{4w^{2}} -
    \frac{w^{2}}{2(\sigma_{p}^{2} + \sigma_{q}^{2})} \left((\Delta
    r_{\parallel} + l)^{2} + \Delta \vec{r}_{\perp}^{2}\right)\right),
\end{multline*}
allows for use of the Poisson summation formula
\begin{equation}
  \sum_{n = -\infty}^{\infty} f(na) = \frac{1}{a} \sum_{n =
    -\infty}^{\infty} \tilde{f}\left(\frac{2\pi n}{a}\right)
\end{equation}
where
\begin{multline*}
    \tilde{f}\left(\frac{2\pi n}{a}\right) = \int_{-\infty}^{\infty}
    e^{-2\pi inl/a} f(l)~\mathrm{d}l = \\ \frac{1}{2\pi w}
    \exp\Bigg(\frac{\theta^{2}}{4} - \frac{\theta^{2}}{4w^{2}}
    + \im (k - 2\pi n/a)\Delta r_{\parallel} \\-
    \frac{\sigma_{p}^{2} + \sigma_{q}^{2}}{2w^{2}} (k - 2\pi
    n/a)^{2} - \frac{w^{2} \Delta
      \vec{r}_{\perp}^{2}}{2(\sigma_{p}^{2} + \sigma_{q}^{2})}\Bigg)
\end{multline*}
is the Fourier transform with respect to the coordinates along the
periodic axis. Using the facts that $\vec{b} = 2\pi\vec{a}/a$ and
$\vec{g} = n\vec{b}$, and that $(k - 2\pi n/a)\Delta r_{\parallel} =
(\vec{k} - \vec{g}) \cdot (\vec{r}_{p} - \vec{r}_{q})$ by definition,
it follows that the integral over $w$,
\begin{multline*}
    \frac{|\vec{b}|}{(2\pi)^{2}} \sum_{\vec{g}}
    \exp(\theta^{2} / 4 + \im(\vec{k} - \vec{g}) \cdot
    (\vec{r}_{p} - \vec{r}_{q})) \times \\ \int_{0}^{\nu} w^{-1}
    \exp(-\rho_{\perp}^{2} w^{2} -
    \frac{\eta^{2}}{4w^{2}})~\mathrm{d}w
\end{multline*}
can be written in terms of the reciprocal lattice sum, having defined
$\eta^{2} = \theta^{2} + 2(\sigma_{p}^{2} + \sigma_{q}^{2})|\vec{k} -
\vec{g}|^{2}$ and $\rho_{\perp} = \frac{|\Delta
  \vec{r}_{\perp}|}{\sqrt{2(\sigma_{p}^{2} + \sigma_{q}^{2})}}$. One
further variable substitution $y = \frac{\nu^{2}}{w^{2}}$ and an
expansion of the exponential term involving $\rho_{\perp}$ in terms of
its Taylor series finally yields the long-range contribution for a
1D-periodic system along an arbitrary axis of periodicity, given by:
\begin{widetext}
\begin{equation*}
    G^{(0)\mathrm{LR}}_{\vec{k}pi,qj} =
    \frac{|\vec{b}|}{8\pi^{2}} (\partial_{r_{pi}} \partial_{r_{pj}} -
    (\xi/c)^{2} \delta_{ij}) \times \sum_{\vec{g}}
    \left(\exp(\theta^{2} / 4 + \im(\vec{k} - \vec{g}) \cdot
    (\vec{r}_{p} - \vec{r}_{q})) \times \sum_{s = 0}^{\infty}
    \frac{(-1)^{s}}{s!} (\nu\rho_{\perp})^{2s} E_{s+1} (\eta^{2} /
    (4\nu^{2}))\right)
\end{equation*}
\end{widetext}
in terms of the exponential integral functions
$E_{1}(x) = \int_{x}^{\infty} t^{-1} e^{-t}~\mathrm{d}t$ and
$E_{s + 1}(x) = s^{-1} (e^{-x} - xE_{s} (x))$, which are closely
related but not identical to incomplete gamma and hypergeometric
functions~\cite{Abramowitz1964}.

For a 2D-periodic system, the lattice vectors and reciprocal lattice
vectors can be defined to lie in a plane with orthonormal vectors
$\vec{e}_{1}$ \& $\vec{e}_{2}$, with vector $\vec{e}_{\perp}$ lying
normal to the plane. This allows for writing
$\vec{R} = R_{1} \vec{e}_{1} + R_{2} \vec{e}_{2}$,
$\vec{g} = g_{1} \vec{e}_{1} + g_{2} \vec{e}_{2}$, and
$\vec{k} = k_{1} \vec{e}_{1} + k_{2} \vec{e}_{2}$, regardless of
lattice geometry, and
$\vec{r}_{p} - \vec{r}_{q} = \Delta r_{1} \vec{e}_{1} + \Delta r_{2}
\vec{e}_{2} + \Delta r_{\perp} \vec{e}_{\perp}$. This means
$\vec{k} \cdot \vec{R} = k_{1} R_{1} + k_{2} R_{2}$ and
$|\vec{r}_{p} + \vec{R} - \vec{r}_{q}|^{2} = (\Delta r_{1} +
R_{1})^{2} + (\Delta r_{2} + R_{2})^{2} + (\Delta
r_{\perp})^{2}$. Once again, from the integral over $w$, the function
\begin{multline*}
    f(l_{1} \vec{e}_{1} + l_{2} \vec{e}_{2}) =
    \frac{\exp(\theta^{2} / 4 - \im(k_{1} l_{1} + k_{2}
      l_{2}))}{2\pi^{3/2} \sqrt{2(\sigma_{p}^{2} + \sigma_{q}^{2})}}
    \times \\ \exp\Bigg(-\frac{\theta^{2}}{4w^{2}} -
    \frac{w^{2}}{2(\sigma_{p}^{2} + \sigma_{q}^{2})} ((\Delta r_{1} +
      R_{1})^{2} \\ + (\Delta r_{2} + R_{2})^{2} + (\Delta
      r_{\perp})^{2})\Bigg)
\end{multline*}
can be used in the Poisson summation formula
\begin{equation}
  \sum_{\vec{R}} f(\vec{R}) = \frac{1}{A_{\mathrm{uc}}} \sum_{\vec{g}}
  \tilde{f}(\vec{g}),
\end{equation}
where the Fourier transform,
\begin{multline*}
    \tilde{f}(g_{1} \vec{e}_{1} + g_{2} \vec{e}_{2}) = \\
    \int_{-\infty}^{\infty} \int_{-\infty}^{\infty} e^{-\im(g_{1}
      l_{1} + g_{2} l_{2})} f(l_{1} \vec{e}_{1} + l_{2}
    \vec{e}_{2})~\mathrm{d}l_{2}~\mathrm{d}l_{1} 
    \\ =\sqrt{\frac{2(\sigma_{p}^{2} + \sigma_{q}^{2})}{\pi}}
    \frac{\exp(\theta^{2} / 4 + \im(\vec{k} + \vec{g})\cdot
      (\vec{r}_{p} - \vec{r}_{q}))}{2w^{2}} \\ 
    \times \exp\left(-\rho_{\perp}^{2} w^{2} -
    \frac{\eta^{2}}{4w^{2}}\right)
\end{multline*}
is written in terms of $\eta$ as above and $\rho_{\perp} =
\frac{\Delta r_{\perp}}{\sqrt{2(\sigma_{p}^{2} +
    \sigma_{q}^{2})}}$. Performing the integration over $w$ finally
yields
\begin{multline*}
    G^{(0)\mathrm{LR}}_{\vec{k}pi,qj} =
    \frac{|\vec{b}_{1} \times \vec{b}_{2}|}{16\pi^{2}}
    (\partial_{r_{pi}} \partial_{r_{pj}} - (\xi/c)^{2} \delta_{ij})
    \times \\ \sum_{\vec{g}} \eta^{-1} \sqrt{2(\sigma_{p}^{2} +
      \sigma_{q}^{2})}\exp\left[\theta^{2} / 4 + \im(\vec{k} +
      \vec{g}) \cdot (\vec{r}_{p} - \vec{r}_{q})\right]
    \\ \times\Big[e^{-\eta\rho_{\perp}} \erfc\left(\frac{\eta}{2\nu}
      - \nu\rho_{\perp}\right) + e^{\eta\rho_{\perp}}
      \erfc\left(\frac{\eta}{2\nu} + \nu\rho_{\perp}\right)\Big]
\end{multline*}
for a 2D-periodic geometry parallel to an arbitrary plane.

In principle, the infinite set of real lattice vectors $\vec{R}$ and
reciprocal lattice vectors $\vec{g}$ must be used for the above
summations. In practice, however, these sums are fast-converging
allowing for truncation after a relatively small number of vectors
$\vec{R}$ and $\vec{g}$, provided an appropriate choice of the Ewald
parameter $\kappa$. The optimal value of this parameter strongly
depends on the separation $|\vec{r}_{p} - \vec{r}_{q}|$, frequency
$\omega$, and effective Gaussian width $\sqrt{2(\sigma_{p}^{2} +
  \sigma_{q}^{2})}$, and the last among those in particular depends
heavily on the material properties of the body in addition to the
geometry; a full convergence analysis is beyond the scope of this
work.

\subsection{vdW interactions and RHT between parallel graphene sheets} \label{sec:graphene}
In our past work~\cite{VenkataramSCIADV2019}, we have considered only
graphene in the RMB framework as it has proved to numerically work
well with the aforementioned Ewald summation procedure; by contrast,
hexagonal boron nitride does not seem to produce such good convergence
properties in practice, and we have not tried other extended periodic
media using our atomistic description in RMB. We also note, as we have
discussed in detail in our previous work~\cite{VenkataramSCIADV2019},
that the atomistic treatment of graphene in RMB ignores the
electromagnetic effects of the interplay between delocalized electrons
and phonons, so our use of graphene in this paper is meant merely to
qualitatively illustrate salient behaviors in fluctuational EM
interactions and to show the convergence and power of the RMB
framework, not to provide high-precision quantitative results to
compare with other theories. As we have already considered the
interactions between a graphene sheet and a parallel PEC
plane~\cite{VenkataramSCIADV2019}, we now consider the interactions
between two parallel graphene sheets in vacuum separated by distance
$d$.

\begin{figure}[t!]
  \centering
  \includegraphics[width=0.95\columnwidth]{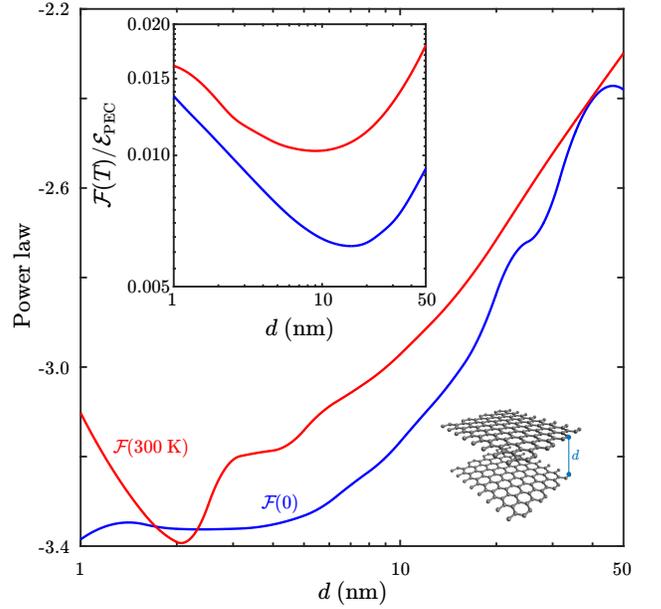}
  \caption{\textbf{vdW interaction between parallel graphene sheets in
      vacuum}. vdW interaction power laws $\frac{\partial
      \ln(\mathcal{F(T)})}{\partial \ln(d)}$ between two parallel
    graphene sheets at separation $d$ at zero (blue) and room (red)
    temperatures. Inset: ratio of the free energy $\mathcal{F}(T)$ to
    the zero-temperature PEC planar interaction energy
    $\mathcal{E}_{\mathrm{PEC}}$.}
  \label{fig:2grapheneVACvdW}
\end{figure}

\begin{figure}[t!]
  \centering
  \includegraphics[width=0.95\columnwidth]{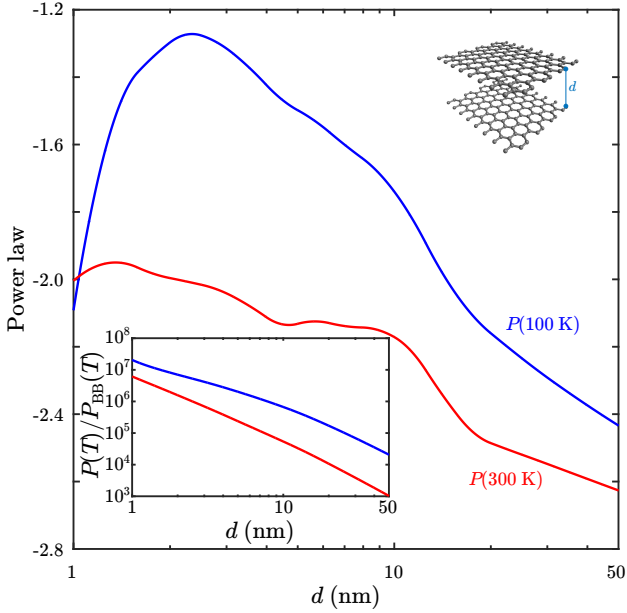}
  \caption{\textbf{RHT between parallel graphene
      sheets in vacuum}. RHT power laws
    $\frac{\partial \ln(\mathcal{P(T)})}{\partial \ln(d)}$ between two
    parallel graphene sheets at separation $d$ at low (blue) and room
    (red) temperatures. Inset: ratio of the exchanged power $P(T)$ to
    the blackbody emission power $P_{\mathrm{BB}}(T)$ at the same
    temperature.}
  \label{fig:2grapheneVACRHT}
\end{figure}

For this system, we consider the vdW interaction free energies
in~\figref{2grapheneVACvdW} at zero temperature $T = 0$ and room
temperature $T = 300~\mathrm{K}$; numerical difficulties in this
system preclude consideration of separations outside of the range $d
\in [1~\mathrm{nm}, 50~\mathrm{nm}]$. The power laws $\frac{\partial
  \ln|\mathcal{F}(T)|}{\partial \ln(d)}$ at both temperatures show
significant deviations from the conventional pairwise prediction of
$-4$ as well as the predictions in the nonretarded random phase
approximation of $-3$~\cite{DobsonJPCM2012}. In particular, both
increase from values more negative than $-3$ at small $d$ to those
less negative than $-3$ at larger $d$, and the room temperature power
law in particular exhibits more sensitivity to $d$ at for $d <
5~\mathrm{nm}$. These behaviors are because the static Gaussian widths
for graphene near the center of the Brillouin zone, which is where the
integrand is dominant, are around $3~\mathrm{nm}$, so the Gaussian
basis functions overlap at such small $d$; the vdW free energy
integrand is more sensitive to static effects at room temperature than
at zero temperature, so the power law is more complicated at room
temperature. Meanwhile, at larger $d$, electromagnetic retardation
interplays with the more complicated material dispersion due to
phonons, leading in the case of two graphene sheets in vacuum to an
initial \emph{increase} rather than a decrease in the power law. All
of these effects are qualitatively very similar to those seen for two
parallel long carbyne wires in vacuum~\cite{VenkataramSCIADV2019},
with quantitative differences arising in the behaviors of these two
carbon allotropes due to greater dimensionality and the lack of finite
size effects in the case of graphene. Additionally, as exemplified in
the ratio of $\mathcal{F}(T)$ to the corresponding zero-temperature
interaction energy of two PEC planes $\mathcal{E}_{\mathrm{PEC}} =
\frac{\pi^{2} \hbar cA}{720d^{3}}$, the free energies themselves are
smooth and monotonic functions of $d$, though the nonmonotonic
behavior of the ratio is exemplified in the behavior of the
corresponding power laws.

We also consider the RHT powers
in~\figref{2grapheneVACRHT} at low temperature $T = 100~\mathrm{K}$
and room temperature $T = 300~\mathrm{K}$ for one of the sheets, where
the other sheet is consistently assumed to be maintained at zero
temperature. The power laws $\frac{\partial \ln|P(T)|}{\partial
  \ln(d)}$ deviate significantly from the prediction of $-4$ by a
pairwise summation of near-field RHT for two
sheets, in both cases behaving nonmonotonically and remaining less
negative than $-3$ in the range of separations of interest due to the
confluence of factors involving the overlap of Gaussian basis
functions particularly for $d < 5~\mathrm{nm}$ and the complicated
interplay of geometry, material dispersion, and electromagnetic
retardation at all separations. That said, the room temperature power
law remains more negative than the low temperature power law due to
the existence of resonances in $\Phi$ at higher frequency that depend
more strongly on separation and are not exponentially suppressed as
they would be at low temperature. The room temperature power law in
particular behaves qualitatively very similarly to that for radiative
heat transfer between two fullerenes in
vacuum~\cite{VenkataramPRL2018}, with quantitative differences again
arising in the behaviors of these two carbon allotropes due to greater
dimensionality and the lack of finite size effects in the case of
graphene; for the system of two graphene sheets, numerical
difficulties again preclude consideration of separations outside of
the range $d \in [1~\mathrm{nm}, 50~\mathrm{nm}]$, but we expect based
on the very similar results of fullerene that as $d$ drops below
$1~\mathrm{nm}$, the power law for decreasing $d$ would continue to
drop toward a local minimum and then sharply increase and essentially
saturate near zero, corresponding to a saturation rather than a
divergence of the RHT power itself with decreasing
$d$ due to the strongly nonlocal material response of graphene as
captured in the atomistic model used in the RMB
framework. Furthermore, the RHT powers themselves
monotonically decay with increasing separation and are significantly
larger than the corresponding blackbody emission powers at each
temperature, though the normalized power is larger at low temperature
largely because the corresponding blackbody emission power is so much
less there than at room temperature.

\section{SIE formulation of interactions among molecules and
  macroscopic bodies} \label{sec:SIE}

We turn to the SIE formulation of Maxwell's equations for the special
case of macroscopic bodies defined by sharp boundaries between regions
where the permittivity is local and homogeneous. This allows for
writing scattering quantities involving macroscopic bodies in terms of
\emph{surface} DOFs and the homogeneous Green's functions on each side
of a surface, rather than volumetric DOFs and associated
susceptibilities. In brief, rather than solving a discretization of
Maxwell's equations in differential or integral form in the full
volume of a body, we instead assign fictitious electric and magnetic
currents to boundaries between permittivity regions and solve for them
by enforcing continuity of the tangential electric and magnetic fields
across each boundary, so that the fields radiated by the fictitious
currents are the scattered fields accounting for multiple scattering
within and between bodies. Note that magnetic surface currents are
needed even for bodies with vanishing magnetic susceptibility, as the
fictitious surface currents are simply the tangential components of
the total fields. The use of surface DOFs already provides a drastic
reduction in computational complexity over methods that use volumetric
DOFs; while the macroscopic surface basis functions
$\{\ket{\vec{b}_{\beta}}\}$ may be spectral or other arbitrary
functions, particular computational gains can be realized via
localized basis functions, such as Rao--Wilton--Glisson (RWG) basis
functions, where the ability to heterogeneously mesh a surface allows
for treatment of general macroscopic surface shapes with arbitrary
features. Below, we define the SIE operators and provide formulas for
the vdW interactions and RHT among molecules and macroscopic objects
in this framework; we do not demonstrate any particular computational
implementation of these formulas, leaving that for future work.

Application of the general formulas using the SIE method requires
appropriate modifications and operator substitutions. In particular,
given a collection of macroscopic objects labeled by the index $n$,
the DOFs are defined on their corresponding surfaces, with
interactions mediated by the exterior vacuum Green's functions
$\GG^{(0)}$ (by our assumption, though the exterior medium could in
principle be a different nontrivial permittivity) and within the
macroscopic body interiors $\GG^{(0,n)}$; having assumed that
the macroscopic bodies are made of homogeneous, local, isotropic
susceptibilities, we clarify that $\GG^{(0,n)}$ is the
\emph{homogeneous} Maxwell Green's function corresponding to the bulk
material constituting macroscopic body $n$, as if its boundaries
didn't exist. We further assume for the purposes of these derivations
that the macroscopic bodies have distinct surfaces and are not
embedded in each other, though the SIE formulation is general enough
to allow for relaxation of those
assumptions~\cite{RodriguezPRB2013}. General scattering problems then
obtained via a SIE scattering operator whose inverse given by
$-\WW^{-1}_\mathrm{mac} = \GG^{(0)} + \sum_{n} \mathbb{S}_n
\GG^{(0,n)} \mathbb{S}_n$~\cite{ReidPRA2013, RodriguezPRB2013}, such
that the scattering Green's function outside of the collection of
macroscopic bodies is $\GG^{(0)} \WW_{\mathrm{mac}} \GG^{(0)}$; here,
$\mathbb{S}_{n}$ is a projection operator onto the \emph{surface}
(rather than volumetric) DOFs of macroscopic body $n$. For the
purposes of vdW interactions as well as thermal emission or heat
transfer only among molecules, the macroscopic bodies only affect the
EM field scattering properties, so the replacements
$\TT_{\mathrm{mac}} \to \WW_{\mathrm{mac}}$ (and analogously
$\TT_{\mathrm{mac}\infty} \to \WW_{\mathrm{mac}\infty}$ for vdW
interactions) are sufficient when evaluating~\eqref{vdWfreeenergy}
and~\eqref{heattransferonlymolormac} in conjunction
with~\eqref{blockVandT}
.

In situations where one seeks to compute energy exchange between a
collection of molecules and a macroscopic body, as may be useful for
localized heating of a molecule by a AFM tip~\cite{CuiNATURE2019}, it
is incumbent to perform additional simplifications (beyond the
substitution $\TT_{\mathrm{mac}} \to \WW_{\mathrm{mac}}$). This is
because the macroscopic DOFs are only defined at their surfaces,
without any reference to volumetric degrees of freedom, so the SIE
formulation leads more naturally to a definition of heat transfer in
terms of the Poynting flux through the surface of a given macroscopic
body $n$ due to fluctuating volumetric polarization sources in
molecule $m$. It is useful to start with the result of first
performing the aforementioned substitution along with $\PP_{n} \to
\mathbb{S}_{n}$ into~\eqref{heattransferbothmolandmac}:
\begin{equation*}
  \begin{split}
    \Phi^{(m)}_{n} = -4~\trace\Big[&\asym(\VV_{m}^{-1\dagger}) \times
      \\ &\PP_{m} (\TT_{\mathrm{mol}}^{-1\dagger} - \GG^{(0)\dagger}
      \WW_{\mathrm{mac}}^{\dagger} \GG^{(0)\dagger})^{-1}
      \GG^{(0)\dagger} \WW_{\mathrm{mac}}^{\dagger} \times
      \\ &\asym(\mathbb{S}_{n} (\WW_{\mathrm{mac}}^{-1} + \GG^{(0)}))
      \times \\ &\WW_{\mathrm{mac}} \GG^{(0)}(\TT_{\mathrm{mol}}^{-1}
      - \GG^{(0)} \WW_{\mathrm{mac}} \GG^{(0)})^{-1} \PP_{m}\Big]
  \end{split}
\end{equation*}
This expression can be further rewritten to obtain a formula that is
conceptually and technically similar to previously derived formulas
for heat transfer between macroscopic
bodies~\cite{RodriguezPRB2013}. First, the combination
$\WW_{\mathrm{mac}}^{-1} + \GG^{(0)}$ is block diagonal in the space
of macroscopic bodies, such that $\mathbb{S}_{n}
(\WW_{\mathrm{mac}}^{-1} + \GG^{(0)}) = -\mathbb{S}_{n}
\mathbb{G}^{(0,n)} \mathbb{S}_{n}$. Hence, the above expression can be
rewritten as:
\begin{equation}
  \begin{split}
    \Phi^{(m)}_{n} = 4~\trace\Big[&\asym(\VV_{m}^{-1\dagger}) \times
      \\ &\PP_{m} (\TT_{\mathrm{mol}}^{-1\dagger} - \GG^{(0)\dagger}
      \WW_{\mathrm{mac}}^{\dagger} \GG^{(0)\dagger})^{-1}
      \GG^{(0)\dagger} \WW_{\mathrm{mac}}^{\dagger} \times
      \\ &\mathbb{S}_{n} \asym(\mathbb{G}^{(0,n)}) \mathbb{S}_{n}
      \times \\ &\WW_{\mathrm{mac}} \GG^{(0)} (\TT_{\mathrm{mol}}^{-1}
      - \GG^{(0)} \WW_{\mathrm{mac}} \GG^{(0)})^{-1} \PP_{m}\Big]
  \end{split}
\end{equation}
Next, if the blockwise inversion to evaluate $\TT$
from~\eqref{blockVandT} is performed accounting for the identity
$(\WW_{\mathrm{mac}}^{-1} - \GG^{(0)}
\TT_{\mathrm{mol}} \GG^{(0)})^{-1} \GG^{(0)}
\TT_{\mathrm{mol}} = \WW_{\mathrm{mac}} \GG^{(0)}
(\TT_{\mathrm{mol}}^{-1} - \GG^{(0)}
\WW_{\mathrm{mac}} \GG^{(0)})^{-1}$, then the heat
transfer between a molecule $m$ and a macroscopic body $n$ can be
written as:
\begin{equation}
  \begin{split}
    \Phi^{(m)}_{n} = 4~\trace\Big[&\asym(\VV_{m}^{-1\dagger}) \times \\ &\PP_{m}
      \TT_{\mathrm{mol}}^{\dagger} \GG^{(0)\dagger}
      (\WW_{\mathrm{mac}}^{-1\dagger} - \GG^{(0)\dagger}
      \TT_{\mathrm{mol}}^{\dagger} \GG^{(0)\dagger})^{-1} \times
      \\ &\mathbb{S}_{n} \asym(\mathbb{G}^{(0,n)}) \mathbb{S}_{n} \times \\
      &(\WW_{\mathrm{mac}}^{-1} - \GG^{(0)} \TT_{\mathrm{mol}}
      \GG^{(0)})^{-1} \GG^{(0)} \TT_{\mathrm{mol}} \PP_{m}\Big].
  \end{split}
\end{equation}
Finally, we may define a modified SIE operator
\begin{equation} \label{eq:Wmacmoddefinition}
  -\WW_{\mathrm{mac}}'^{-1} = \GG^{(0)} + \GG^{(0)} \TT_{\mathrm{mol}}
  \GG^{(0)} + \sum_{n} \mathbb{S}_{n} \mathbb{G}^{(0,n)}
  \mathbb{S}_{n}
\end{equation}
in analogy to~\eqref{Tmolmoddefinition}, as an effective SIE operator
where the exterior medium is no longer vacuum but encodes the
scattering properties (to infinite order) of the \emph{molecules} as a
background medium. This allows for writing the radiative energy
exchange between a molecule $m$ and a macroscopic body $n$ can be
written as:
\begin{equation}
  \begin{split}
    \Phi^{(m)}_{n} = 4~\trace\Big[&\asym(\VV_{m}^{-1\dagger}) \PP_{m}
      \TT_{\mathrm{mol}}^{\dagger} \GG^{(0)\dagger}
      \WW_{\mathrm{mac}}'^{\dagger} \times \\ &\mathbb{S}_{n}
      \asym(\mathbb{G}^{(0,n)}) \mathbb{S}_{n} \times
      \\ &\WW_{\mathrm{mac}}' \GG^{(0)} \TT_{\mathrm{mol}}
      \PP_{m}\Big]
  \end{split}
\end{equation}
in a more compact way. Conceptually, this formula describes the
energy transfer as a Poynting flux from volumetric sources in molecule
$m$, whose correlations are proportional to
$\asym(\VV_{m}^{-1\dagger})$, through the surface of
macroscopic body $n$ via
$\mathbb{S}_{n} \GG^{(0,n)} \mathbb{S}_{n}$, where scattering
between all of the molecules and macroscopic bodies is accounted to
all orders via the combination of $\WW_{\mathrm{mac}}'$ and
$\TT_{\mathrm{mol}}$.

It is exactly this substitution in~\eqref{Wmacmoddefinition} that
further allows for computing the heat transfer among macroscopic
bodies in the presence of molecules in the SIE framework. In such a
case, the energy flow from the surface of one macroscopic body (due to
fluctuations in its interior) through the surface of another is
desired, with the molecules simply modifying the scattering properties
of the medium exterior to the macroscopic objects. This makes the heat
transfer between macroscopic bodies $m$ \& $n$
\begin{equation}
  \begin{split}
    \Phi^{(m)}_{n} = 4~\trace\Big[&\asym(\GG^{(0,m)}) \mathbb{S}_{m}
      \WW_{\mathrm{mac}}'^{\dagger} \mathbb{S}_{n} \times
      \\ &\asym(\GG^{(0,n)}) \mathbb{S}_{n} \WW_{\mathrm{mac}}'
      \mathbb{S}_{m}\Big]
  \end{split}
\end{equation}
the same as that in~\cite{RodriguezPRB2013},
using~\eqref{Wmacmoddefinition} in the presence of the molecular
bodies.

In all of these formulas, the molecular DOFs are expressed in terms of
the Gaussian basis functions $\ket{\vec{f}_{pi}}$ as usual, while the
macroscopic DOFs are expressed in terms of basis functions denoted
$\ket{\vec{b}_{\beta}}$: the latter may in principle be either
spectral or localized basis functions, but localized RWG basis
functions are preferred for convergence in arbitrary macroscopic
geometries that do not have a high degree of translational or
rotational symmetry. This means that the expression of
$\TT_{\mathrm{mol}}$ in terms of $\ket{\vec{f}_{pi}}$, along with the
matrix elements $\bracket{\vec{f}_{pi}|\GG^{(0)} \vec{f}_{qj}}$,
$\bracket{\vec{b}_{\beta}|\GG^{(0)} \vec{b}_{\beta'}}$, and
$\bracket{\vec{b}_{\beta}|\GG^{(0,n)} \vec{b}_{\beta'}}$ (for a
macroscopic body labeled $n$ of a given homogeneous susceptibility)
are needed: routines to compute these matrix elements have already
been implemented, the former two in our new code and the latter two in
the SCUFF-EM boundary element solver~\cite{SCUFF}. However, on top of
this, the matrix elements $\bracket{\vec{f}_{pi}|\GG^{(0)}
  \vec{b}_{\beta}}$ need to be computed as well: this has not yet been
implemented, but may be done through appropriate conjunction of the
SCUFF-EM code with our code as both are open source software. We do
note that just as for molecular bodies above a PEC plane, these
derivations assume that the molecular basis functions can be
associated purely with the space external to the macroscopic bodies,
which is not exactly true given that Gaussian basis functions do not
have compact support, and this assumption becomes somewhat more
questionable when the center of a basis function is less than one
Gaussian width away from the boundary of a macroscopic body. That
said, this approach should still qualitatively capture the effects of
screening on interactions between molecules and macroscopic bodies
even at such short separations, and is an improvement on our previous
approximation of molecular basis functions as point dipoles in their
interactions with macroscopic bodies~\cite{VenkataramPRL2017} (which
was only justifiable in the absence of phonons so that the Gaussian
widths were much smaller than the considered separations between
molecules and macroscopic bodies).

\section{Concluding Remarks} \label{sec:conclusions}
The RMB formulation of fluctuational electrodynamics makes clear that
``molecular'' and ``macroscopic'' bodies can be treated on the same
footing, given appropriate atomistic or continuum descriptions of
each. It allows atomistic descriptions of material bodies based on
coupled effective electronic and nuclear oscillators, accounting for
short-range electronic correlations and phonons whose properties are
obtained from ab-initio density functional theory calculations, and is
in principle compatible with arbitrary continuum descriptions of
material response as well. It can in principle be extended to account
for material bodies treated with continuum response theories when such
bodies have arbitrary shapes (beyond simple planar structures).
Furthermore, the power of this formulation lies in the
\emph{analytical} formulas for the electromagnetic interaction matrix
elements of material bodies treated atomistically using the
aforementioned oscillator model, sidestepping questions of convergence
common to finite-volume or discrete-dipole computational techniques by
assigning Gaussian basis functions created from material response
properties obtained within the RMB framework itself.


There are several shortcomings and open questions that require further
attention. Chief among them is that the atomistic oscillator model is
physically accurate only for insulating or weakly conducting system,
and is less appropriate for strongly metallic or semimetallic systems
where electron delocalization effects are more visible in conjunction
with phonons and long-range electromagnetic interactions. This has
been discussed in detail in our prior
work~\cite{VenkataramSCIADV2019}, particularly concerning how the RMB
framework can capture the salient geometric and phononic properties of
graphene and related atomically-thin materials like hexagonal BN,
which will be similar, but cannot capture the inherent electron
delocalization in graphene that is absent from polar dielectric media
like hexagonal BN. Related to this, in extended media where the
effects of phonon and electron delocalization would be most relevant
if present, the Ewald summation procedure applied to Gaussian basis
functions constructed from the susceptibility within each unit cell is
not guaranteed in practice to yield numerically well-behaved results:
for example, proper convergence is obtained for infinite sheets of
graphene, but not for infinite sheets of hexagonal BN.

Even for compact molecules, the widths of the Gaussian basis functions
encode information about the anisotropy of the molecule as a whole but
are not themselves anisotropic for each atomic basis function. It
remains to be seen for a broader variety of molecules interacting at
the mesoscale the extent to which such a change in the basis functions
may make a difference, but that is beyond the scope of this
work. Furthermore, for compact molecules and extended atom-scale
structures, DFT calculations may yield effective internuclear coupling
matrices $K_{\mathrm{I}}$ that go far beyond nearest neighbors, but
numerical convergence of such long-range couplings is not always
guaranteed in practice; therefore, some care must be taken in plugging
those matrices into code built on the RMB framework, and it may be
necessary to restrict couplings to nearest or next-nearest neighbors.

The extension to include continuum bodies of arbitrary geometries has
yet to be computationally implemented in practice. That is beyond the
scope of this work, but we imagine this to be relatively
straightforward. However, the extent to which the possible overlap of
relatively large atomistic Gaussian basis functions with hard
boundaries of continuum bodies may degrade the accuracy of predictions
in the RMB framework remains to be tested through direct comparisons
with relevant past~\cite{WagnerNATURE2014, LoskillJRSI2013,
  TsoiACSNANO2014} and future experiments; these would be the ultimate
tests of the reliability and \emph{raison d'\^{e}tre} of our RMB
framework.

Finally, we note that while the measurements of near-field RHT between
metallic tips and substrates at nanometric gaps~\cite{KimNATURE2015,
  KloppstechNATURE2017, CuiNATURE2017, ChiloyanNATURE2015} can be
modeled using the RMB framework if electrons and phonons are localized
within each body to preclude the possibility of conduction between
bodies, the current RMB framework is unable to model total heat
transfer when both radiative and conductive processes between bodies
are relevant, particularly in the extreme near-field. Extending the
RMB framework and associated code to handle such situations is the
subject of an upcoming manuscript.

Further testing in diverse combinations of molecular and macroscopic
bodies will doubtless yield more questions, so addressing all of these
issues will be the subject of many future works. We anticipate that
other researchers may be able to make use of our code both to model
mesoscale FED phenomena in specific systems and to answer some of
these broader outstanding questions.

\emph{Acknowledgments.}---This work was supported by the National
Science Foundation under Grants No. DMR-1454836, DMR 1420541, and DGE
1148900, the Cornell Center for Materials Research MRSEC (award no.
DMR 1719875), the Defense Advanced Research Projects Agency (DARPA)
under Agreement No. HR00112090011, and the FNR CORE project
QUANTION. The views, opinions, and/or findings expressed herein are
those of the authors and should not be interpreted as representing the
official views or policies of any institution.

\appendix
\section{Computational details}



Each of the inverse T-operators $\TT_{\mathrm{mol(mac)}}^{-1}$ can
further be written blockwise, with the diagonal blocks representing
the inverse T-operator of a given molecular or macroscopic body and
the off-diagonal blocks encoding the EM fields propagated between the
corresponding pair of bodies. Thus, in any basis representation, the
diagonal blocks of the Green's function and inverse T-operator matrix
representations are independent of the relative separations or
orientations of the molecular or macroscopic bodies, so if EM
interaction quantities are desired for multiple separations or
orientations, these diagonal blocks need to only be computed once per
frequency; only the off-diagonal blocks need to be recomputed for
every change in separation or orientation between a given pair of
bodies.

\section{Glossary of terms} \label{sec:glossary}

Here, in Table~\ref{tab:glossary}, we present a glossary of terms
relating quantities and their conventional notations in quantum
chemistry literature versus continuum FED literature.

\begin{widetext}
\begin{table}[t] 
\centering
\begin{tabular}{|c|c|c|}
  \hline
  Quantum chemistry & Continuum FED & Relationship \\
  \hline \hline
  $\chi^{(0)}(\omega, \vec{x}, \vec{x}')$ & $V_{ij}(\omega, \vec{x},
  \vec{x}')$ & $\chi^{(0)} (\omega, \vec{x}, \vec{x}') = \sum_{i,j}
  \partial_{i} \partial_{j} V_{ij}(\omega, \vec{x}, \vec{x}')$
  \\ \hline  
  $v(\vec{x}, \vec{x}') = \frac{1}{4\pi|\vec{x} - \vec{x}'|}$ &
  $G^{(0)}_{ij}(\omega, \vec{x}, \vec{x}') = (\partial_{i}
  \partial_{j} + (\omega/c)^{2} \delta_{ij})\frac{e^{\im \omega|\vec{x} -
    \vec{x}'|/c}}{4\pi|\vec{x} - \vec{x}'|}$ & $G^{(0)}_{ij}(0,
  \vec{x}, \vec{x}') = \partial_{i} \partial_{j} v(\vec{x}, \vec{x}')$
  \\ \hline
  $\chi_{\mathrm{RPA}}(\omega, \vec{x}, \vec{x}')$: & $T_{ij}(\omega,
  \vec{x}, \vec{x}')$: & $\chi_{\mathrm{RPA}}(\omega, \vec{x},
  \vec{x}') = $ \\
  $\chi_{\mathrm{RPA}} = \chi^{(0)} + \chi^{(0)} v\chi_{\mathrm{RPA}}$
  & $\TT = \VV + \VV\GG^{(0)}\TT$ & $\sum_{i,j} \partial_{i}
  \partial_{j} T^{\mathrm{nonret}}_{ij} (\omega, \vec{x}, \vec{x}')$
  \\ \hline

\end{tabular}
\caption{Glossary: note that $\TT^{\mathrm{nonret}}$ is computed at
  each $\omega$ such that $\VV$ is evaluated at that $\omega$ but
  $\GG^{(0)}$ is evaluated at zero frequency.}
\label{tab:glossary}
\end{table}
\end{widetext}

\bibliography{unifiedmesoscopiclongpaper}

\end{document}